%% file: main.tex
\documentclass[letterpaper]{article} % DO NOT CHANGE THIS
\usepackage{aaai2027} % DO NOT CHANGE THIS

\usepackage[hyphens]{url} % DO NOT CHANGE THIS
\usepackage{graphicx} % DO NOT CHANGE THIS
\usepackage{natbib} % DO NOT CHANGE THIS AND DO NOT ADD OPTIONS
\usepackage{caption} % DO NOT CHANGE THIS AND DO NOT ADD OPTIONS
\usepackage{amsmath}
\usepackage{amssymb}
\usepackage{algorithm}
\usepackage{algorithmic}
\usepackage{booktabs}
\usepackage{adjustbox}
\usepackage{booktabs,multirow}
\usepackage[table]{xcolor}
\definecolor{tableheader}{RGB}{221,229,237}
\usepackage{multirow}

\title{From Inaudible Inputs to Model Failures: Low-Frequency Safety Risks in LALMs}

\author{
 \textbf{Yuanhe Zhang\textsuperscript{1,$^\star$}}, 
 \textbf{Weiliu Wang\textsuperscript{1,$^\star$}}, 
 \textbf{Jie Ren\textsuperscript{1}}, 
 \textbf{Liang Lin\textsuperscript{2},}
 \textbf{Zhenhong Zhou\textsuperscript{3},} 
 \\
 \textbf{Haoran Gao\textsuperscript{4},}
 \textbf{Kun Wang\textsuperscript{3},} 
 \textbf{Chen Li\textsuperscript{5},}
 \textbf{Li Sun\textsuperscript{1},} 
 \textbf{Sen Su\textsuperscript{1, 6, $^\dagger$}} 
\\ \textsuperscript{\rm 1}Beijing University of Posts and Telecommunications \\
\textsuperscript{\rm 2}Institute of Information Engineering, Chinese Academy of Sciences \quad
\\ \textsuperscript{\rm 3}Nanyang Technological University \quad
\textsuperscript{\rm 4}JIUTIAN Research\\
\textsuperscript{\rm 5}Tencent ARC Lab
\textsuperscript{\rm 6}Chongqing University of Posts and Telecommunications
\\ \{charmes-zhang, susen\}@bupt.edu.cn
}
\affiliations{}

\begin{document}

\maketitle

\begingroup
\renewcommand\thefootnote{}\footnotemark
\renewcommand\thefootnote{}\footnotemark
\footnotetext{\hspace{-1.8em}$\star$ indicates equal contribution.\\
\hspace{-1.5em}$\dagger$ indicates corresponding author.}
\endgroup
% https://anonymous.4open.science/r/Low-Frequency-Lockout-80C7

\begin{abstract}
\input{sections/abstract}
\end{abstract}

% Optional anonymous links. Keep this block between the abstract and main body.
% \begin{links}
%     \link{Code}{https://anonymous.example/code}
%     \link{Datasets}{https://anonymous.example/data}
% \end{links}

\input{Figures/main_fig}

\input{sections/introduction}

\input{sections/related_work}
\input{sections/method}

\input{sections/experiments}
\input{sections/analysis}

\input{sections/limitations}

\input{Table/table_analyse}

\input{sections/conclusion}

\bibliography{main}

\clearpage
\appendix
\setcounter{secnumdepth}{1}
\input{sections/appendix}

% Uncomment if AAAI 2027 requires the reproducibility checklist in the paper.
% \input{ReproducibilityChecklist.tex}

\end{document}

%% file: sections/abstract.tex
Large audio-language models (LALMs) have demonstrated strong capabilities in understanding diverse audio inputs.
This diversity includes low-frequency signals that are inaudible to humans but can still enter the model and influence its generation.
However, the practical impact of such low-frequency inputs on LALMs remains largely unexplored.
In this paper, we propose \textit{Intermittent Low-Frequency Lockout} (ILL), an inaudible red teaming method that evaluates this risk using a universal waveform template in a black box setting.
ILL uses \textit{Sentence Attention Scale Estimation} to determine active intervals and \textit{Frequency Confusion Transfer} to construct a low-frequency waveform with continuous phase from corpus spectral variation.
To mitigate this risk, we propose \textit{Distributional Requery Guard} (DRG) to detect low-frequency distribution shifts and conditionally request a second recording for semantic recovery.
Across six LALMs and multiple audio understanding tasks, ILL reduces accuracy by up to 67 percentage points while receiving a mean human audibility rating of 1.33, close to 1.17 for clean audio; DRG raises mean attacked accuracy from 28.5\% to 46.1\% after clean reacquisition.
These findings identify a previously overlooked safety risk for LALMs and provide a foundation for future research on robust audio understanding.
% Our code is available at \url{https://anonymous.4open.science/r/Low-Frequency-Lockout-80C7}.

%% file: Figures/main_fig.tex
\begin{figure*}[t]
\centering
\includegraphics[width=\textwidth]{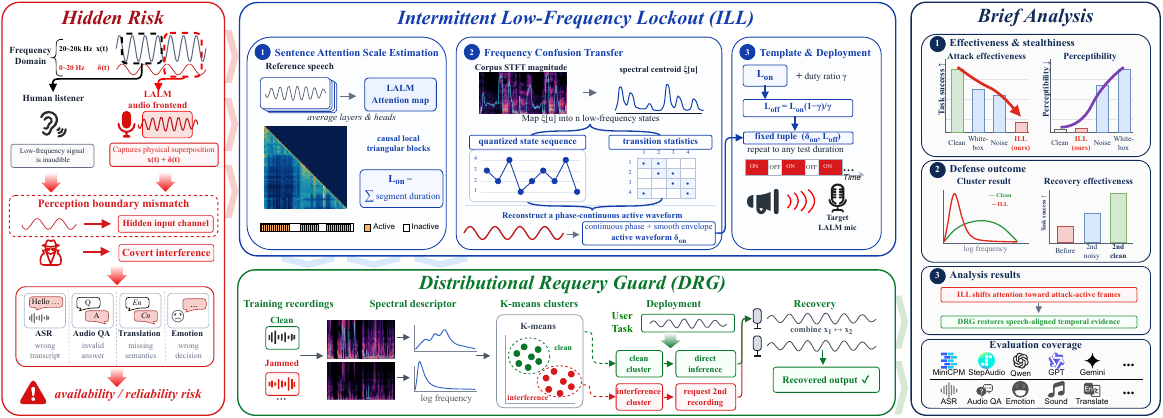}
\caption{Overview of the low-frequency perception gap, the construction of ILL, the DRG detection pipeline, and the evaluation.}
\label{Fig:main}
\vspace{-9pt}
\end{figure*}

%% file: sections/introduction.tex
\section{Introduction}
\label{sec:introduction}

Recent large audio-language models (LALMs) combine acoustic perception with language reasoning to understand speech, environmental sounds, and music within a unified interface~\citep{chu2023qwen}. Their audio encoders transform sampled waveforms into model-side acoustic representations~\citep{panayotov2015librispeech}, so the resulting input surface is not defined by human perception alone. A sufficiently low-frequency signal may therefore enter an LALM's audio frontend while remaining imperceptible to its user, creating a potential mismatch between the acoustic evidence available to the model and to the person interacting with it.

Prior work has shown that malicious or corrupted audio inputs can compromise both task performance and safety in audio-driven models~\citep{sadasivan2026attacker,yin2026focus,song2026audio}. Adversarial audio can manipulate model behavior or disrupt task execution~\citep{carlini2018audio,sadasivan2026attacker}, while acoustic corruption can degrade audio understanding and weaken safety performance
~\citep{yin2026focus,song2026audio}. Together, these findings expose the sensitivity of LALMs to audio input that may not be salient to users but can still affect model behavior.
Low-frequency interference may intensify these risks because users cannot perceive the interfering signal. However, the safety risk posed by low-frequency interference to LALMs remains underexplored.

In this paper, we propose \textit{Intermittent Low-Frequency Lockout} (ILL), an inaudible red teaming method that evaluates this risk using a universal waveform in a black box setting. ILL combines \textit{Sentence Attention Scale Estimation}, which derives active intervals from multiscale changes in audio attention, with \textit{Frequency Confusion Transfer}, which converts corpus spectral variation into a continuous-phase low-frequency state sequence. The resulting intermittent waveform remains fixed across test recordings.
We further propose \textit{Distributional Requery Guard} (DRG), which detects low-frequency distribution shifts and conditionally requests a second recording for joint semantic recovery.

We evaluate ILL across six LALMs and multiple audio understanding tasks. ILL reduces accuracy by up to 67 percentage points, while its audible noise ratio remains at 0.06--0.08\% and its mean human audibility rating is 1.33, close to 1.17 for clean audio. DRG achieves detection F1 scores of 89.69--99.00\% and improves defense performance by 17.6 percentage points, raising the six-model mean accuracy from 28.5\% to 46.1\%. The same requery mechanism also recovers useful semantic evidence under other noise perturbations and attack methods, indicating transfer beyond ILL. Our internal analyses further suggest that ILL's performance degradation is associated with reduced attention to acoustic evidence and lower output confidence.

Our contributions are as follows:
\begin{itemize}
    \item We identify the hidden risk posed by inaudible low-frequency inputs to LALMs and introduce ILL, a black box red teaming method for validating this risk.
    \item We develop DRG, which detects low-frequency distribution shifts and mitigates the resulting risk through conditional requery and joint semantic recovery.
    \item We evaluate ILL and DRG across six LALMs and multiple audio tasks, and link low-frequency interference to reduced acoustic attention and output confidence.
\end{itemize}

%% file: sections/related_work.tex
\section{Related Work}
\label{sec:related_work}

\subsection{Large Audio-Language Models}
\label{sec:related_work_lalms}
Large audio-language models (LALMs) connect acoustic representations with language-model reasoning to support speech interaction and general audio understanding \cite{zhang2023speechgpt,rubenstein2023audiopalm,chu2023qwen,chu2024qwen2}. By exposing open-ended language generation to continuous acoustic inputs, these models extend the capabilities of text-only interfaces while introducing audio input surfaces that can influence model behavior.

\subsection{Attack Risks in Large Audio-Language Models}
\label{sec:related_work_adversarial_audio}
Before LALMs, adversarial audio research showed that speech systems could be controlled through imperceptible or disguised acoustic commands~\cite{carlini2016hidden,zhang2017dolphinattack,carlini2018audio,yuan2018commandersong,yan2020surfingattack}. 
With LALMs, this attack surface extends from command recognition to open-ended behavior and task completion. Existing red teaming mainly examines safety failures induced by malicious or misleading audio inputs \cite{peri2024speechguard,yang2025audio,yang2025can,chen2026audiojailbreak}. Research also broadens this evaluation to naturalistic speech and signal-level manipulations \cite{yu2026now,song2026audio, lin2026hidden}. Another line studies availability failures in which audio disrupts task execution or redirects model behavior under digital and over-the-air settings \cite{hou2025evaluating,ma2025universal,sadasivan2026attacker}. However, the availability risk posed by a fixed low-frequency physical signal that is independent of the utterance remains underexplored.

\subsection{LALM Safety and Audio-Side Defenses}
\label{sec:related_work_lalm_robustness}
\label{sec:related_work_audio_defenses}
Existing LALM defenses primarily restore safe behavior through attack-aware checks or corrections around model inference \cite{yang2025speech,djanibekov2025spirit,hsu2025reducing, zhang2026see}. A complementary direction adapts conventional audio denoising to LALMs through plug-and-play enhancement before inference \cite{yin2026focus}. Although these approaches address semantic attacks and general acoustic corruption, detecting stealthy low-frequency interference remains insufficiently studied.

%% file: sections/method.tex
\section{Method}
\label{sec:method}
% Figure~\ref{Fig:main} summarizes the studied perception gap, the construction and deployment of ILL, and the DRG mitigation pipeline.
% This section presents \textit{Intermittent Low-Frequency Lockout} (ILL), a universal black box attack for red teaming large audio language models (LALMs), together with a preliminary mitigation. We first specify the physical attack setting and adversary capabilities. We then construct the ILL attack in two stages: \textit{Sentence Attention Scale Estimation} derives the duration and spacing of active attack intervals, and \textit{Frequency Confusion Transfer} constructs the low-frequency variation within each interval and synthesizes the final intermittent waveform. We finally introduce \textit{Distributional Requery Guard}.

\subsection{Threat Model}
\label{sec:method_threat_model}

As illustrated in Figure~\ref{Fig:main}, we consider an inaudible attack in which an adversary emits a low-frequency waveform while a user speaks to a target LALM; the two signals superpose in air and are captured as a composite waveform by the target microphone. Measurements show that small commercial loudspeakers and electret microphones retain a nonzero response over 0.5--20~Hz, while smartphone microphones remain measurable across the infrasonic band in non-isolated settings~\citep{jeng2011response,asmar2018digital}. As with high-power radio-frequency counter-drone jammers, practical reach depends on source power, propagation loss, and receiver response. The adversary constructs a universal waveform offline using a reference speech corpus and a fixed reference LALM, without accessing or querying the target or adapting the waveform to individual test utterances, enabling the same waveform to attack different utterances and target LALMs. A fuller analysis of the acoustic-environment--digital-input chain is provided in Appendix~\ref{sec:appendix_physical_chain}.

\subsection{Intermittent Low-Frequency Lockout (ILL) Attack}
\label{sec:method_ill_attack}

\subsubsection{Sentence Attention Scale Estimation}
\label{sec:method_sentence_attention}

Let training data $\mathcal{D}$ contain $M$ recordings, and let $x\in\mathcal{D}$ denote one recording with duration $T_x$. The reference LALM $\mathcal{R}$ converts audio at a model-specific rate of $\rho\in\mathbb{N}^{+}$ tokens per second and produces $F$ audio tokens for $x$. We compute the mean corpus duration $\bar{T}=M^{-1}\sum_{x\in\mathcal{D}}T_x$ for the duration normalization in \textit{Frequency Confusion Transfer}.

We perform a forward pass of each $x$ through $\mathcal{R}$, which contains $L$ layers and $H$ attention heads per layer. Let $\mathbf{A}^{l,h}$ denote the self-attention matrix from layer $l$ and head $h$. 
We average these matrices as $\bar{\mathbf{A}}=(LH)^{-1}\sum_{l=1}^{L}\sum_{h=1}^{H}\mathbf{A}^{l,h}$ and define $\mathbf{B}\in\mathbb{R}^{F\times F}$ as the submatrix of $\bar{\mathbf{A}}$ indexed by the $F$ audio tokens along both dimensions.
We then normalize each valid row of this causal audio attention matrix:
\begin{equation}
    \mathbf{P}[p,q]=
    \frac{\mathbf{B}[p,q]}
    {\sum_{a=0}^{p}\mathbf{B}[p,a]+\varepsilon},
    \qquad 0\leq q\leq p<F.
    \label{eq:normalized_attention}
\end{equation}
Here $\varepsilon>0$ prevents division by zero. The resulting $\mathbf{P}[\cdot]$ retains the local lower-triangular attention blocks associated with continuous semantic processing and exposes weak attention across potential transitions.

We detect these transitions at $\mathcal{W}=\{\rho,5\rho,10\rho\}$. For a candidate boundary $b\in\{1,\ldots,F-1\}$ and a scale $w\in\mathcal{W}$, let $\mathcal{L}_{b,w}=\{\max(0,b-w),\ldots,b-1\}$ and $\mathcal{R}_{b,w}=\{b,\ldots,\min(F-1,b+w-1)\}$ index the tokens immediately to its left and right. We obtain the cross-boundary and within-segment attention masses through matrix slicing:
\begin{equation}
    \begin{aligned}
        C_w(b)&=\left\|\mathbf{P}[\mathcal{R}_{b,w},\mathcal{L}_{b,w}]\right\|_{1,1},\\
        I_w(b)&=\left\|\operatorname{tril}\!\left(\mathbf{P}[\mathcal{R}_{b,w},\mathcal{R}_{b,w}]\right)\right\|_{1,1}.
    \end{aligned}
    \label{eq:boundary_attention_mass}
\end{equation}
The norm $\|\cdot\|_{1,1}$ sums the absolute values of all entries, and $\operatorname{tril}(\cdot)$ retains the lower triangle. For each $w$, we smooth the relative attention mass over $\mathcal{N}_w(b)=\{u:\max(1,b-w)\leq u\leq\min(F-1,b+w)\}$ and average the results across scales to compute the final boundary score:
\begin{equation}
    g(b)=
    \frac{1}{|\mathcal{W}|}
    \sum_{w\in\mathcal{W}}
    \frac{1}{|\mathcal{N}_w(b)|}
    \sum_{u\in\mathcal{N}_w(b)}
    \frac{I_w(u)}
    {I_w(u)+C_w(u)}.
    \label{eq:smoothed_multiscale_score}
\end{equation}
We calculate the median $m$ and median absolute deviation $d$ over $\{g(b)\}_{b=1}^{F-1}$, then retain local maxima satisfying $g(b)\geq m+d$. After adding the two endpoints, we obtain $0=b_0<\cdots<b_Q=F$, where $Q$ denotes the number of resulting intervals. Each adjacent pair yields duration $(b_{q+1}-b_q)/\rho$. After repeating this procedure for all recordings in $\mathcal{D}$, we set $L_{\mathrm{on}}$ to the median of the resulting durations.
This corpus-derived duration aligns each active interval with a typical span of continuous semantic attention. Given a duty ratio $\gamma\in(0,1]$, we compute the inactive duration and cycle length:
\begin{equation}
    \begin{aligned}
        L_{\mathrm{off}}=L_{\mathrm{on}}\frac{1-\gamma}{\gamma},
        T_{\mathrm{cyc}}=L_{\mathrm{on}}+L_{\mathrm{off}}.
    \end{aligned}
    \label{eq:intermittent_schedule}
\end{equation}
The schedule places an active interval at the beginning of the recording and repeats it every $T_{\mathrm{cyc}}$ until the recording ends.

\subsubsection{Frequency Confusion Transfer}
\label{sec:method_audio_semantic_transfer}

To disturb the target LALM's perception of speech semantics, this module transfers corpus-level acoustic variation into the frequency trajectory of each active interval. We divide the low-frequency range $[f_{\min},f_{\max}]$ uniformly into $n\geq2$ states and assign state $j\in\{1,\ldots,n\}$ the following frequency:
\begin{equation}
    f_j=f_{\min}+\frac{j-1}{n-1}(f_{\max}-f_{\min}).
    \label{eq:low_frequency_states}
\end{equation}
The parameter $n$ controls the granularity of this state space.

For each $x\in\mathcal{D}$, we apply the short-time Fourier transform~\cite{allen1977unified} to capture how its local spectral content changes over time. Let $\mathbf{X}_x\in\mathbb{C}^{U_x\times V}$ denote the resulting spectrogram, where $U_x$ is the number of time frames in $x$ and $V$ is the number of frequency bins. Let $\nu_v$ denote the frequency represented by bin $v$. We summarize the local spectrum of frame $u$ through its spectral centroid:
\begin{equation}
    \xi_x[u]
    =
    \frac{\sum_{v=1}^{V}\nu_v|\mathbf{X}_x[u,v]|}
    {\sum_{v=1}^{V}|\mathbf{X}_x[u,v]|+\varepsilon},
    \qquad 1\leq u\leq U_x.
    \label{eq:spectral_centroid}
\end{equation}

The corpus extrema $\xi_{\min}=\min_{x\in\mathcal{D},\,1\leq u\leq U_x}\xi_x[u]$ and $\xi_{\max}=\max_{x\in\mathcal{D},\,1\leq u\leq U_x}\xi_x[u]$ define a shared quantization range across all recordings.
Each centroid is mapped to one of the $n$ states as $y_x[u]=\min\!\left(n,\,1+\left\lfloor n(\xi_x[u]-\xi_{\min})/(\xi_{\max}-\xi_{\min}+\varepsilon)\right\rfloor\right)$ for $1\leq u\leq U_x$.
The resulting sequence $\mathbf{y}_x=(y_x[1],\ldots,y_x[U_x])$ preserves the temporal order of the local spectral variation in $x$.
We aggregate $\{\mathbf{y}_x:x\in\mathcal{D}\}$ to estimate the corpus-level state statistics. Let $\mathbb{I}[\cdot]$ denote the indicator function. For each state $j$, $c_j$ counts its occurrences across all sequences, while $c_{jk}$ counts adjacent transitions from state $j$ to state $k$, $j,k\in\{1,\ldots,n\}$. We compute these counts as:
\begin{equation}
    \begin{aligned}
        c_j
        &=
        \sum_{x\in\mathcal{D}}
        \sum_{u=1}^{U_x}
        \mathbb{I}[y_x[u]=j],\\
        c_{jk}
        &=
        \sum_{x\in\mathcal{D}}
        \sum_{u=1}^{U_x-1}
        \mathbb{I}[y_x[u]=j,\,y_x[u+1]=k].
    \end{aligned}
    \label{eq:state_counts}
\end{equation}
Let $N_{\mathrm{obs}}=\sum_{x\in\mathcal{D}}U_x$ denote the total number of state observations and let $c_j^{\rightarrow}=\sum_{k=1}^{n}c_{jk}$ denote the number of transitions leaving state $j$. We normalize these corpus counts to obtain the  transition matrix $\mathbf{Q}[\cdot,\cdot]$:
\begin{equation}
    \begin{aligned}
        \pi_j
        =
        \frac{c_j}{N_{\mathrm{obs}}},
        \mathbf{Q}[j,k]
        =
        \frac{c_{jk}+\alpha}
        {c_j^{\rightarrow}+n\alpha}.
    \end{aligned}
    \label{eq:transition_matrix}
\end{equation}
The coefficient $\alpha>0$ provides Laplace smoothing for unobserved transitions. Duration normalization allocates $K=\operatorname{round}(T_{\mathrm{cyc}}/\bar{T})+1$ equal-length subsegments to each active interval. Using the previously computed $\mathbf{Q}[\cdot,\cdot]$, we decode the most probable state sequence $\mathbf{s}^{\star}=(s_1,\ldots,s_K)$:
\begin{equation}
    \mathbf{s}^\star=
    \underset{\substack{
    s_1,\ldots,s_K\in\{1,\ldots,n\}\\
    s_k\neq s_{k-1},\ 2\leq k\leq K
    }}{\arg\max}
    \left[
    \sum_{k=2}^{K}\log\mathbf{Q}[s_{k-1},s_k]
    \right].
    \label{eq:state_path}
\end{equation}
Dynamic programming yields this path while enforcing a state change between adjacent subsegments. Each subsegment lasts $\Delta=L_{\mathrm{on}}/K$ and uses frequency $f_{s_k}$. To model the noise intensity received by the target microphone, we use amplitude $\beta$ together with a smooth envelope $e(t)$ whose endpoints vanish, then synthesize the active waveform with continuous phase:
\begin{equation}
    \begin{aligned}
        &\varphi(t)=f_{s_k},\quad
        (k-1)\Delta\leq t<k\Delta, k=1,\ldots,K,\\
        &\delta_{\mathrm{on}}(t)=\beta e(t)
        \sin\!\left(2\pi\int_{0}^{t}\varphi(u)\,du\right),
        0\leq t<L_{\mathrm{on}}.
    \end{aligned}
    \label{eq:active_waveform}
\end{equation}
For a test recording, let $\tau(t)=t\bmod T_{\mathrm{cyc}}$. We repeat the active waveform according to the schedule from \textit{Sentence Attention Scale Estimation} to produce the full perturbation:
\begin{equation}
    \delta^\star(t)=
    \begin{cases}
        \delta_{\mathrm{on}}(\tau(t)),
        & 0\leq\tau(t)<L_{\mathrm{on}},\\
        0,
        & L_{\mathrm{on}}\leq\tau(t)<T_{\mathrm{cyc}},
    \end{cases}.
    \label{eq:intermittent_waveform}
\end{equation}
The attacker emits $\delta^\star$ as a standalone acoustic signal rather than digitally mixing it with the test recording. In simulation, $\beta$ controls the noise intensity received by the target microphone, while the acoustic environment performs the physical superposition. The offline-constructed tuple $\delta^\star(t)$ forms the universal intermittent template and remains fixed across test recordings and target LALMs.

\subsection{Distributional Requery Guard}
\label{sec:method_distributional_requery_guard}

\textit{Distributional Requery Guard} (DRG) is a plug-in input filter that detects low-frequency distribution shifts before LALM inference and requests independent acoustic evidence only for flagged inputs. 
Offline, we construct a detector training set $\mathcal{D}_{\mathrm{def}}=\mathcal{D}_{\mathrm{cle}}\cup\mathcal{D}_{\mathrm{jam}}$, where $\mathcal{D}_{\mathrm{cle}}$ contains clean training recordings and $\mathcal{D}_{\mathrm{jam}}$ contains training recordings with low-frequency interference. For each defense training recording $x_{\mathrm{def}}\in\mathcal{D}_{\mathrm{def}}$, we apply the short-time Fourier transform specified in Section~\ref{sec:method_audio_semantic_transfer}, sum the resulting magnitudes over time, and apply $\ell_1$ normalization across frequency bins. This operation produces a spectral distribution descriptor $\mathbf{d}_x\in\mathbb{R}^{V}$, whose component $\mathbf{d}_x[v]$ represents the relative magnitude in frequency bin $v$.
The normalization reduces sensitivity to overall recording amplitude while preserving the spectral distribution used for detection.

We cluster the training descriptors into two groups with K-means. Let $\mathcal{G}_1$ and $\mathcal{G}_2$ denote the resulting sets of training recordings, and let $\boldsymbol{\mu}_k\in\mathbb{R}^{V}$ denote the centroid of cluster $\mathcal{G}_k$. Both fitting and assignment use Euclidean distance. To determine which cluster represents interference, we compute its mean descriptor mass within the same low-frequency range $[f_{\min},f_{\max}]$ used to construct ILL:
\begin{equation}
        r_k=\frac{1}{|\mathcal{G}_k|}
        \sum_{x\in\mathcal{G}_k}
        \sum_{v:\,\nu_v\in[f_{\min},f_{\max}]}\mathbf{d}_x[v],
        \quad k\in\{1,2\}.
    \label{eq:defense_cluster_label}
\end{equation}
We label the cluster with the larger mass as the interference cluster, $k_{\mathrm{jam}}=\arg\max_{k\in\{1,2\}}r_k$, and label the remaining cluster as $k_{\mathrm{cle}}=3-k_{\mathrm{jam}}$. This procedure depends only on the input spectra and does not query the target LALM.

At deployment, DRG extracts the descriptor of the user's initial recording $x^{(1)}$ and assigns it to the nearest centroid. A recording assigned to $k_{\mathrm{cle}}$ follows the standard inference path. For a recording assigned to $k_{\mathrm{jam}}$, the system requests a second recording $x^{(2)}$ of the same utterance and supplies both recordings to the target LALM with an instruction to compare them and answer only from content that is consistent and intelligible across the pair. The second recording therefore provides independent acoustic evidence without discarding usable content from the initial recording. Given $\mathbf{d}_x$, DRG adds only two $V$-dimensional distance evaluations, yielding $\mathcal{O}(V)$ detection time without a neural forward pass or target-LALM query; Appendix~\ref{sec:appendix_drg_complexity} provides the full complexity analysis.

%% file: sections/experiments.tex
\section{Experiments}
\label{sec:experiments}

\input{Table/table_attack_task_accuracy_mini}

\input{Table/table_invisibility_subjective_evaluation}

% We evaluate the effectiveness and stealthiness of \textit{Intermittent Low-Frequency Lockout} (ILL), followed by defense and benign-utility tests.

\subsection{Experimental Setup}
\label{sec:experiments_setup}

\textbf{Models.}
We evaluate four open-weight LALMs: MiniCPM-o~4.5 (MiniCPM)~\citep{cui2026minicpm}, StepAudio2 Mini (StepAudio2)~\citep{wu2025step}, Qwen2.5-Omni (Qwen2.5)~\citep{Xu2025Qwen25OmniTR}, and Qwen3-Omni (Qwen3)~\citep{xu2025qwen3}. We also test two closed-source LALMs, GPT-audio-mini (GPT) and Gemini-3.5-flash (Gemini). We construct ILL once with Qwen2.5-Omni as the reference model and transfer the fixed waveform to all other targets.
% [DETAIL NEEDED: Pin model revisions/API snapshots and report decoding, compute, and software settings.]

\textbf{Tasks and Datasets.}
We use MMAU's speech, sound, and music categories~\citep{sakshi2025mmau} for audio question answering, LibriSpeech~\citep{panayotov2015librispeech} for speech recognition, English-to-Chinese CoVoST~2~\citep{wang2021covost} for speech translation, and RAVDESS~\citep{livingstone2018ryerson} for emotion classification. We sample 100 test examples per dataset and construct ILL using another disjoint set of 100 examples.

\textbf{Attack Baselines.}
We compare ILL with four attacks. \textit{Gaussian} substitutes random Gaussian noise~\citep{franceschi2018robustness} under ILL's duty cycle. \textit{PNL} selects one common interference recording from each of three 115 Noise categories: natural, mechanical, and human~\citep{ko2017study}.
\textit{Audio-Adv} is a targeted white-box attack that induces chosen DeepSpeech transcriptions~\citep{carlini2018audio}. \textit{Whisper} is a universal acoustic attack that suppresses Whisper transcription~\citep{raina2024muting}. Clean and attacked conditions use paired source audio and matched budgets where applicable.
% [DETAIL NEEDED: Specify the common budget, Gaussian placement, PNL recordings, and baseline implementations.]

\textbf{Defense Baselines.}
We compare DRG with direct low-frequency suppression (LF-Suppression), Deep Feature Loss (DFL) speech denoising~\citep{germain2018speech}, and MMSE-STSA enhancement~\citep{ephraim1984speech}.

\textbf{Metrics and Evaluation Protocol.}
We report accuracy for MMAU and RAVDESS. LibriSpeech uses WER, whereas CoVoST~2 uses BLEU~\citep{papineni2002bleu}. We assess objective stealthiness with the audible noise ratio defined in Section~\ref{sec:experiments_stealthiness}; lower values indicate that less perturbation energy falls in the nominal audible band. Subjective stealthiness uses human ratings from 1 to 7, where lower values indicate lower perceived audibility. Defense detection uses precision, recall, and F1, while recovery uses the corresponding end-task metric after requery.

\subsection{Attack Effectiveness}
\label{sec:experiments_attack_effectiveness}
\input{Figures/fig_attack_bleu_wer_by_model}

Table~\ref{tab:attack_accuracy_mini} and Figure~\ref{Fig:fig_attack_bleu_wer_by_model} show that the fixed ILL waveform degrades all four task types. This effect does not weaken after black-box transfer: the largest RAVDESS reduction, 67 percentage points, occurs on the unseen StepAudio2 target rather than on the Qwen2.5 reference model. The consistent degradation on the other five LALMs therefore indicates that ILL has strong migration capabilities.

ILL does not uniformly outperform every audible-noise baseline, but it achieves comparable and, in some settings, stronger disruption using only the 5--20~Hz band.
It gives the lowest CoVoST~2 BLEU on every target in Figure~\ref{Fig:fig_attack_bleu_wer_by_model} and remains competitive on the accuracy-based tasks.
The key finding is therefore not uniform numerical dominance, but that a low-perceptibility frequency band alone can cause broad degradation across LALMs.
% Conventional noise is stronger in a few individual settings, but nearly all of its energy lies inside the measured band. The advantage of ILL is thus the combination of broad attack effectiveness and low spectral audibility, rather than uniform dominance in every setting.

\subsection{Acoustic Stealthiness}
\label{sec:experiments_stealthiness}

Table~\ref{tab:audibility_score} reports the proportion of perturbation energy that falls in the measured audible band. Human hearing is conventionally characterized over approximately 20~Hz--20~kHz~\citep{Salt_2010}. Because ILL operates at 5--20~Hz, we quantify this spectral leakage as:
\begin{equation}
    \mathrm{ANR}=100\times
    \frac{E_{\mathrm{noise}}(20\,\mathrm{Hz}\text{--}8\,\mathrm{kHz})}
         {E_{\mathrm{noise}}(\text{full band})},
    \label{eq:audible_noise_ratio}
\end{equation}
where lower ANR indicates less perturbation energy in this band. Across the six datasets, ILL has an ANR of only 0.06--0.08\%, whereas all evaluated baselines exceed 98.9\%. ILL therefore achieves its task-level effect with negligible spectral leakage into the measured audible band.

\input{Figures/fig_human_violin}
We complement ANR with 112 complete human response sets covering clean audio and all seven interference methods. Figure~\ref{Fig:fig_human_violin} shows that ILL receives a mean rating of 1.33 and a median of 1, close to the clean condition with a mean of 1.17 and a median of 1. Every comparison method receives a higher mean rating; the closest are Audio-Adv. at 3.75 and Whisper at 3.92. The agreement between the human ratings and ANR supports the low perceptibility of ILL in this evaluation without treating spectral leakage alone as evidence of human imperceptibility.

\subsection{Defense Evaluation}
\label{sec:experiments_defense_evaluation}

\input{Table/table_defense_f1_identification}
DRG first identifies whether the input contains low-frequency interference. As shown in Table~\ref{tab:drg_detection}, the Average training condition yields F1 scores of 89.69--99.00\% across the four evaluation datasets. The result indicates that the spectral distribution provides a reliable trigger for conditional requery.

\input{Figures/fig_defense_stage2_effectiveness}

We then evaluate recovery under two realistic conditions. A clean second recording represents transient interference, whereas a noisy second recording represents persistent interference. As shown in Figure~\ref{Fig:fig_defense_stage2_effectiveness}, DRG raises the six-model mean accuracy from 28.5\% without defense to 46.1\% after clean reacquisition, outperforming the signal-processing baselines. It also remains above the no-defense mean under persistent interference, although the smaller gain confirms that recovery depends on the quality of the second recording.

\input{Table/table_defense_stage2_harmlessness}
\input{Figures/fig_defense_stage2_transferability}

We also test the same recovery procedure on other attack types. Figure~\ref{Fig:fig_defense_stage2_transferability} shows that clean reacquisition is best or tied for best in 19 of 24 settings. This result indicates that DRG's requery mechanism can recover useful semantic evidence beyond the low-frequency interference used in its design.

\subsection{Benign Utility}
\label{sec:experiments_benign_utility}

Table~\ref{tab:clean_drg_comparison} reports a conservative stress test in which DRG is forced to requery every clean input. Ten of the twelve scores remain within 0.03 of standard inference, and DRG exceeds standard inference on two classification results. The utility change is therefore small in most evaluated settings.

%% file: Table/table_attack_task_accuracy_mini.tex
% Format the change relative to the clean score:
% drops up to 10 pp are gray, larger drops are red, and gains are unsigned and green.
\ExplSyntaxOn
\cs_new:Npn \attack_delta:n #1
  {
    \fp_compare:nNnTF {#1} < {0}
      {\textcolor{green!50!black}{\fp_eval:n {round(abs(#1),0)}\%}}
      {
        \fp_compare:nNnTF {#1} > {10}
          {\textcolor{red}{\fp_eval:n {round(#1,0)}\%}}
          {\textcolor{gray}{\fp_eval:n {round(#1,0)}\%}}
      }
  }
\cs_new_eq:NN \attackdelta \attack_delta:n
\ExplSyntaxOff
\newcommand{\score}[2]{%
  \(#1\%\,_{\scriptscriptstyle \attackdelta{#2}}\)%
}
\newcommand{\bestscore}[2]{%
  \(\mathbf{#1\%}\,_{\scriptscriptstyle \mathbf{\attackdelta{#2}}}\)%
}

\begin{table*}[t]
\centering
\caption{Attack results on accuracy-based tasks. MMAU scores are averaged
over its speech, sound, and music categories. Subscripts report the accuracy
change relative to the clean setting in percentage points (pp): gray denotes
drops of at most 10 pp (including no change), red denotes drops greater than
10 pp, and green denotes gains.}
\label{tab:attack_accuracy_mini}

\small
\setlength{\tabcolsep}{3.5pt}
\renewcommand{\arraystretch}{1.15}

\begin{adjustbox}{max width=\textwidth,center}
\begin{tabular}{@{}l|l|c|cccc|c@{}}
\toprule

\rowcolor[HTML]{DDE5ED}
\textbf{Model}\quad\quad\quad\quad
& \textbf{Dataset}\quad\quad\quad\quad\quad
& \quad\quad\textbf{Clean}\quad\quad\quad
& \quad\quad\textbf{Gaussian}\quad\quad\quad
& \quad\quad\textbf{PNL}\quad\quad\quad
& \quad\quad\textbf{Audio-Adv.}\quad\quad\quad
& \quad\quad\textbf{Whisper}\quad\quad\quad
& \quad\quad\textbf{ILL (Our)}\quad\quad\quad \\
\midrule

% ==================== Qwen2.5 ====================
\cellcolor{white}{}
& MMAU
& 73.3\%
& \score{61.3}{12.0}
& \score{60.9}{12.4}
& \score{69.3}{4.0}
& \score{72.3}{1.0}
& \score{61.3}{12.0} \\

\rowcolor[HTML]{F5F7F9}
\cellcolor{white}\multirow{-2}{*}{Qwen2.5}
& RAVDESS
& 47.0\%
& \score{12.0}{35.0}
& \score{11.3}{35.7}
& \score{14.0}{33.0}
& \score{33.0}{14.0}
& \bestscore{8.0}{39.0} \\
\midrule

% ==================== Qwen3 ====================
\cellcolor{white}{}
& MMAU
& 74.7\%
& \score{57.3}{17.3}
& \score{60.6}{14.0}
& \score{69.3}{5.3}
& \score{71.7}{3.0}
& \score{59.3}{15.3} \\

\rowcolor[HTML]{F5F7F9}
\cellcolor{white}\multirow{-2}{*}{Qwen3}
& RAVDESS
& 29.0\%
& \score{12.0}{17.0}
& \score{11.0}{18.0}
& \score{14.0}{15.0}
& \score{30.0}{-1.0}
& \bestscore{8.0}{21.0} \\
\midrule

% ==================== MiniCPM ====================
\cellcolor{white}{}
& MMAU
& 71.7\%
& \score{61.0}{10.7}
& \score{60.9}{10.8}
& \score{69.3}{2.3}
& \score{71.0}{0.7}
& \bestscore{60.3}{11.3} \\

\rowcolor[HTML]{F5F7F9}
\cellcolor{white}\multirow{-2}{*}{MiniCPM}
& RAVDESS
& 18.0\%
& \score{9.0}{9.0}
& \score{9.3}{8.7}
& \score{9.0}{9.0}
& \score{13.0}{5.0}
& \score{10.0}{8.0} \\
\midrule

% ==================== StepAudio ====================
\cellcolor{white}{}
& MMAU
& 70.0\%
& \score{60.7}{9.3}
& \score{58.4}{11.6}
& \score{67.3}{2.7}
& \score{68.0}{2.0}
& \bestscore{56.3}{13.7} \\

\rowcolor[HTML]{F5F7F9}
\cellcolor{white}\multirow{-2}{*}{StepAudio}
& RAVDESS
& 75.0\%
& \score{12.0}{63.0}
& \score{11.3}{63.7}
& \score{32.0}{43.0}
& \score{53.0}{22.0}
& \bestscore{8.0}{67.0} \\
\midrule

% ==================== GPT ====================
\cellcolor{white}{}
& MMAU
& 61.0\%
& \score{51.0}{10.0}
& \score{54.0}{7.0}
& \score{58.3}{2.7}
& \score{60.7}{0.3}
& \bestscore{50.0}{11.0} \\

\rowcolor[HTML]{F5F7F9}
\cellcolor{white}\multirow{-2}{*}{GPT}
& RAVDESS
& 9.0\%
& \score{2.0}{7.0}
& \score{3.3}{5.7}
& \score{2.0}{7.0}
& \score{4.0}{5.0}
& \bestscore{0.0}{9.0} \\
\midrule

% ==================== Gemini ====================
\cellcolor{white}{}
& MMAU
& 73.7\%
& \score{61.7}{12.0}
& \score{58.7}{15.0}
& \score{71.0}{2.7}
& \score{69.7}{4.0}
& \bestscore{58.3}{15.3} \\

\rowcolor[HTML]{F5F7F9}
\cellcolor{white}\multirow{-2}{*}{Gemini}
& RAVDESS
& 40.0\%
& \score{17.0}{23.0}
& \score{13.3}{26.7}
& \score{27.0}{13.0}
& \score{38.0}{2.0}
& \bestscore{12.0}{28.0} \\

\bottomrule
\end{tabular}
\end{adjustbox}

\vspace{-5pt}
\end{table*}

%% file: Table/table_invisibility_subjective_evaluation.tex
\begin{table*}[t]
\centering
\caption{Audible noise ratio (ANR, \%) for clean audio and seven interference methods across four datasets.}
\label{tab:audibility_score}

\small
\setlength{\tabcolsep}{5pt}
\renewcommand{\arraystretch}{1.15}

\begin{adjustbox}{max width=\textwidth,center}
\begin{tabular}{@{}l | c | c c c c c c| c@{}}
\toprule

\rowcolor[HTML]{DDE5ED}
\textbf{Dataset}\quad\quad\quad\quad
& \quad\textbf{Clean}\quad\quad
& \quad\textbf{Gaussian}\quad\quad
& \quad\textbf{PNL (Natural)}\quad\quad
& \quad\textbf{PNL (Machine)}\quad\quad
& \quad\textbf{PNL (Human)}\quad\quad
& \quad\textbf{Audio-Adv.}\quad\quad
& \quad\textbf{Whisper}\quad\quad
& \quad\textbf{ILL (Our)}\quad\quad \\
\midrule

% 第一行：白色
MMAU
& 0.00\%
& 99.63\%
& 100.00\%
& 99.69\%
& 100.00\%
& 98.94\%
& 98.93\%
& \textbf{0.08\%} \\

% 第二行：极浅灰色
\rowcolor[HTML]{F5F7F9}
LibriSpeech
& 0.00\%
& 99.63\%
& 100.00\%
& 99.65\%
& 100.00\%
& 98.94\%
& 98.93\%
& \textbf{0.07\%} \\

% 第三行：白色
CoVoST2
& 0.00\%
& 99.63\%
& 100.00\%
& 99.67\%
& 100.00\%
& 98.94\%
& 98.93\%
& \textbf{0.07\%} \\

% 第四行：极浅灰色
\rowcolor[HTML]{F5F7F9}
RAVDESS
& 0.00\%
& 99.62\%
& 100.00\%
& 99.68\%
& 99.99\%
& 98.94\%
& 98.93\%
& \textbf{0.06\%} \\

\bottomrule
\end{tabular}
\end{adjustbox}

\vspace{-5pt}
\end{table*}

%% file: Figures/fig_attack_bleu_wer_by_model.tex
\begin{figure}[t]
\centering
\includegraphics[width=\columnwidth]{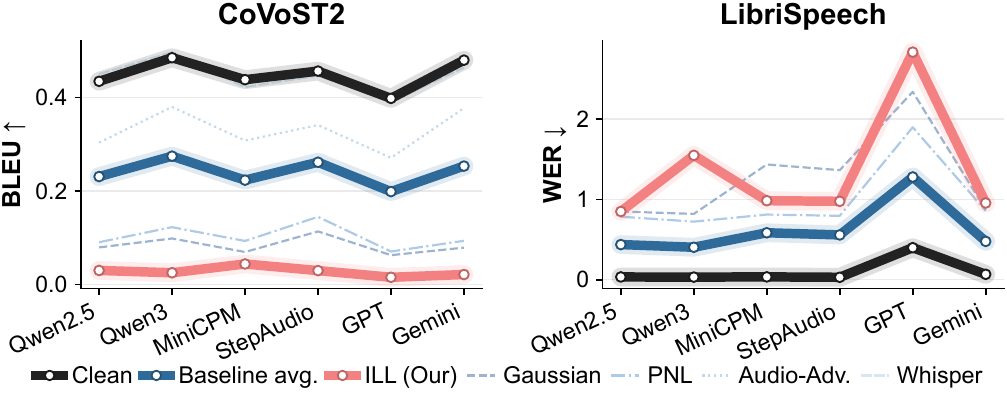} % Reduce the figure size so that it is slightly narrower than the column.
\caption{Attack effectiveness on translation and recognition. Lower BLEU and higher WER indicate stronger disruption.
% Baseline avg. denotes the mean of the four comparison attacks.
}
\label{Fig:fig_attack_bleu_wer_by_model}
\vspace{-9pt}
\end{figure}

%% file: Figures/fig_human_violin.tex
\begin{figure}[t]
\centering
\includegraphics[width=\columnwidth]{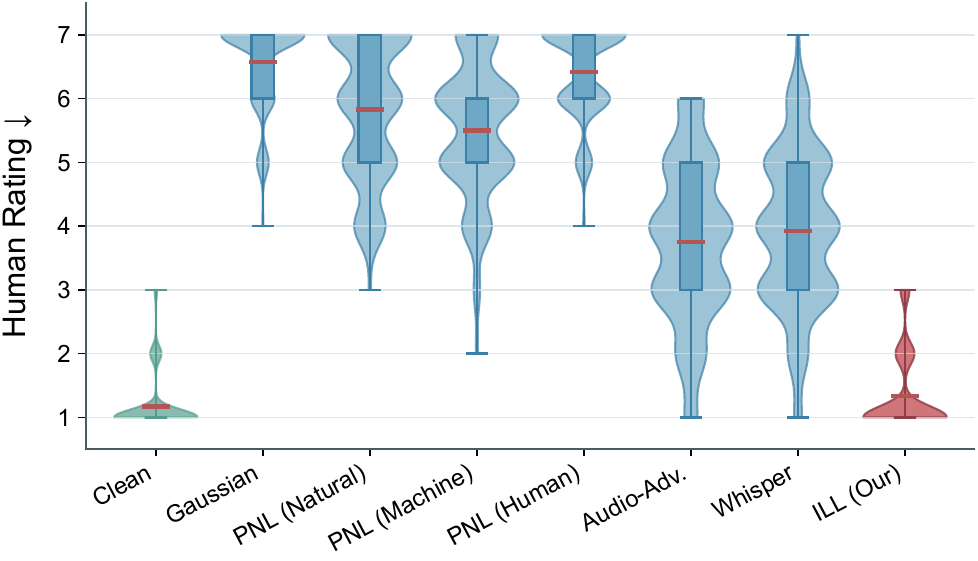}
\caption{Human audibility ratings across 100+ responses.}
\label{Fig:fig_human_violin}
\vspace{-9pt}
\end{figure}

%% file: Table/table_defense_f1_identification.tex
% ==================== 导言区需要的包（确保已引入）
% \usepackage{booktabs}
% \usepackage{xcolor}
% \usepackage{multirow}
% \usepackage{adjustbox}
% \usepackage{stackengine}

% ==================== 正文表格 ====================
\begin{table}[t]
\centering
\caption{Precision, recall, and F1 scores (\%) for DRG interference detection under different detector training datasets.
% Columns identify the evaluation dataset.
}
\label{tab:drg_detection}

\footnotesize
\setlength{\tabcolsep}{2.4pt}
\renewcommand{\arraystretch}{1.15}

\begin{adjustbox}{max width=\columnwidth,center}
\begin{tabular}{@{}ll|cccc@{}}
\toprule

\rowcolor[HTML]{DDE5ED}
\textbf{Training dataset}\quad\quad
&
\textbf{Metric}\quad\quad\quad
&
\textbf{\shortstack[c]{MMAU}}
&
\textbf{\shortstack[c]{LibriSpeech}}
&
\textbf{CoVoST2}
&
\textbf{RAVDESS} \\

\midrule

% ==================== LibriSpeech ====================
\multirow{3}{*}{LibriSpeech}
& Precision & 93.40 & 93.46 & 88.50 & 59.17 \\
& \cellcolor[HTML]{F5F7F9}Recall
& \cellcolor[HTML]{F5F7F9}99.00
& \cellcolor[HTML]{F5F7F9}100.00
& \cellcolor[HTML]{F5F7F9}100.00
& \cellcolor[HTML]{F5F7F9}100.00 \\
& \cellcolor[HTML]{E5E9ED}F1-score
& \cellcolor[HTML]{E5E9ED}96.12
& \cellcolor[HTML]{E5E9ED}\textbf{96.62}
& \cellcolor[HTML]{E5E9ED}93.90
& \cellcolor[HTML]{E5E9ED}74.35 \\

\midrule

% ==================== CoVoST2 ====================
\multirow{3}{*}{CoVoST2}
& Precision & 88.24 & 85.19 & 86.41 & 60.24 \\
& \cellcolor[HTML]{F5F7F9}Recall
& \cellcolor[HTML]{F5F7F9}88.24
& \cellcolor[HTML]{F5F7F9}85.19
& \cellcolor[HTML]{F5F7F9}86.41
& \cellcolor[HTML]{F5F7F9}60.24 \\
& \cellcolor[HTML]{E5E9ED}F1-score
& \cellcolor[HTML]{E5E9ED}\textbf{89.11}
& \cellcolor[HTML]{E5E9ED}88.46
& \cellcolor[HTML]{E5E9ED}87.68
& \cellcolor[HTML]{E5E9ED}75.19 \\

\midrule
% ==================== Average ====================
\multirow{3}{*}{Average}
& Precision & 99.00 & 97.98 & 100.00 & 81.30 \\
& \cellcolor[HTML]{F5F7F9}Recall
& \cellcolor[HTML]{F5F7F9}99.00
& \cellcolor[HTML]{F5F7F9}97.00
& \cellcolor[HTML]{F5F7F9}97.00
& \cellcolor[HTML]{F5F7F9}100.00 \\
& \cellcolor[HTML]{E5E9ED}F1-score
& \cellcolor[HTML]{E5E9ED}\textbf{99.00}
& \cellcolor[HTML]{E5E9ED}97.49
& \cellcolor[HTML]{E5E9ED}98.48
& \cellcolor[HTML]{E5E9ED}89.69 \\

\bottomrule
\end{tabular}
\end{adjustbox}

\end{table}

%% file: Figures/fig_defense_stage2_effectiveness.tex
\begin{figure}[t]
\centering
\includegraphics[width=\columnwidth]{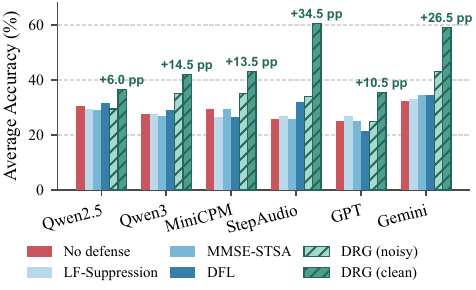} % Reduce the figure size so that it is slightly narrower than the column.
\caption{Defense effectiveness of DRG.}
\label{Fig:fig_defense_stage2_effectiveness}
\vspace{-9pt}
\end{figure}

%% file: Table/table_defense_stage2_harmlessness.tex
% 导言区需要：
% \usepackage{booktabs}
% \usepackage{adjustbox}
% \usepackage[table]{xcolor}
% \usepackage{colortbl}

\begin{table}[t]
\centering
\caption{Benign task performance under standard clean inference and forced DRG requery. 
% Arrows beside each task metric indicate whether higher or lower values are better.
}
\label{tab:clean_drg_comparison}

\footnotesize
\setlength{\tabcolsep}{3pt}
\renewcommand{\arraystretch}{1.15}
\arrayrulecolor{black}

\begin{adjustbox}{max width=\columnwidth,center}
\begin{tabular}{@{}l|cc|cc|cc@{}}
\toprule

% 第一行表头
\rowcolor[HTML]{DDE5ED}
\textbf{Dataset}
& \multicolumn{2}{c|}{\textbf{MiniCPM}}
& \multicolumn{2}{c|}{\textbf{StepAudio}}
& \multicolumn{2}{c}{\textbf{Qwen3}} \\

% 第二行表头
\rowcolor[HTML]{DDE5ED}
&
\quad Clean\quad\quad
&\quad DRG\quad\quad
&\quad Clean\quad\quad
&\quad DRG\quad\quad
& \quad Clean\quad\quad
&\quad DRG \quad\quad\\

\midrule

% 第一行：白色
LibriSpeech$\ _{WER\downarrow}$
& 0.0348
& 0.0374
& 0.0298
& 0.0309
& 0.0320
& 0.0370 \\

% 第二行：浅灰色
\rowcolor[HTML]{F5F7F9}
CoVoST2$\ _{BLEU\uparrow}$
& 0.4382
& 0.4268
& 0.4566
& 0.4455
& 0.4852
& 0.4839 \\

% 第三行：白色
MMAU$\ _{ACC\uparrow}$
& 72.0\%
& 69.0\%
& 66.0\%
& 69.0\%
& 75.0\%
& 62.0\% \\

% 第四行：浅灰色
\rowcolor[HTML]{F5F7F9}
RAVDESS$\ _{ACC\uparrow}$
& 18.0\%
& 21.0\%
& 75.0\%
& 74.0\%
& 29.0\%
& 25.0\% \\

\bottomrule
\end{tabular}
\end{adjustbox}

\end{table}

%% file: Figures/fig_defense_stage2_transferability.tex
\begin{figure}[t]
\centering
\includegraphics[width=\columnwidth]{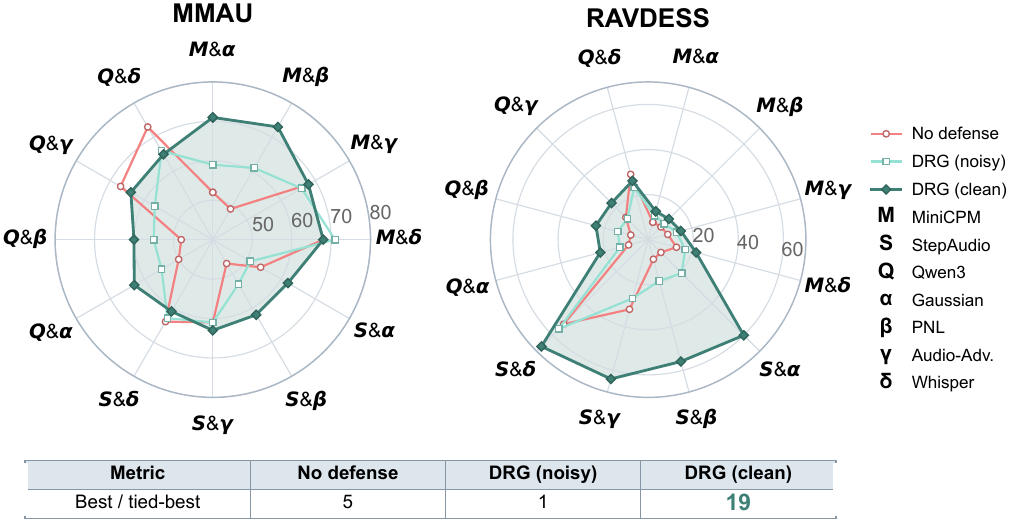} % Reduce the figure size so that it is slightly narrower than the column.
\caption{Recovery under four additional attack types.}
\label{Fig:fig_defense_stage2_transferability}
\vspace{-9pt}
\end{figure}

%% file: sections/analysis.tex
\section{Analysis}
\label{sec:analysis}

This section isolates which design choices make ILL effective and examines model-internal evidence for the observed failures. Unless otherwise stated, controlled ablations use Qwen3-Omni, while mechanistic comparisons retain paired clean, attacked, and defended inputs.

\subsection{Ablation Studies}
\label{sec:analysis_ablations}

\input{Figures/fig_ablation_component_analysis}

Figure~\ref{Fig:fig_ablation_component_analysis} tests whether ILL depends on a particular operating point or on its structured low-frequency construction. Lower task performance indicates a stronger attack in all four panels.

\textbf{Perturbation Amplitude.}
Figure~\ref{Fig:fig_ablation_component_analysis}(a) shows a steep performance decline up to $\beta=4$, followed by saturation. We select this knee point because it captures most of the attack gain without using the largest tested amplitude.

\textbf{Duty Cycle.}
Figure~\ref{Fig:fig_ablation_component_analysis}(b) shows that attack effectiveness improves as the active proportion increases, but the curve changes little beyond $\gamma>70\%$. The selected duty cycle therefore reaches the saturation region while leaving part of the waveform inactive.

\textbf{Attack Duration.}
Figure~\ref{Fig:fig_ablation_component_analysis}(c) reveals a clear duration threshold at the default segment length. Shorter segments are ineffective, whereas extending the segment by one or two orders of magnitude provides no further gain. The default length is therefore sufficient to trigger the observed failure.

\textbf{State-Sequence Construction.}
Figure~\ref{Fig:fig_ablation_component_analysis}(d) shows that ILL causes greater degradation than Gaussian noise, a fixed-frequency signal, and a uniform frequency sweep on both classification and translation. This consistent advantage supports the effectiveness of the proposed state-sequence design.

\subsection{Temporal Segmentation Consistency.}
\input{Table/table_attack_time}
Table~\ref{tab:attention_interval_duration} examines whether \textit{Sentence Attention Scale Estimation} yields comparable active-interval durations across datasets and reference LALMs. All estimates lie between 3.67 and 5.07 seconds, and the largest cross-model spread within a dataset is 0.50 seconds. This consistency supports using a corpus-level attention statistic to determine the ILL schedule rather than relying on model-specific timing.

\subsection{Attention and Representation Shift}
\label{sec:analysis_attention_shift}

Table~\ref{tab:internal_representation} shows that ILL reduces the attention mass assigned to audio from 0.0416 to 0.0373 on MMAU and from 0.0777 to 0.0531 on RAVDESS. DRG partially restores this mass in both cases. This shift is consistent with reduced reliance on acoustic evidence during generation.

The encoded representation retains semantic alignment with the clean input, but that alignment weakens under attack. Cosine similarity falls from 1.0000 to 0.8659 on MMAU and to 0.6309 on RAVDESS, before recovering to approximately 0.96 under DRG. The joint reduction in audio attention and representation similarity is consistent with reduced use of acoustic evidence and a shift in its encoded semantics.

\subsection{Output Confidence Degradation}
\label{sec:analysis_confidence}

ILL also lowers confidence at the decision stage. In Table~\ref{tab:internal_representation}, the probability assigned to the correct answer drops from 0.8050 to 0.1709 on MMAU and from 0.8240 to 0.0179 on RAVDESS. Confidence in the model's selected final answer also decreases on both datasets, showing that the effect is not limited to a change in answer correctness. DRG moves all four confidence values back toward their clean counterparts. Together with the attention and representation shifts, these results are consistent with low-frequency interference changing the model's internal evidence use and downstream decision making.
% [DETAIL NEEDED: Define probability extraction and normalization for multi-token or free-form answers.]

%% file: Figures/fig_ablation_component_analysis.tex
\begin{figure*}[t]
\centering
\includegraphics[width=\textwidth]{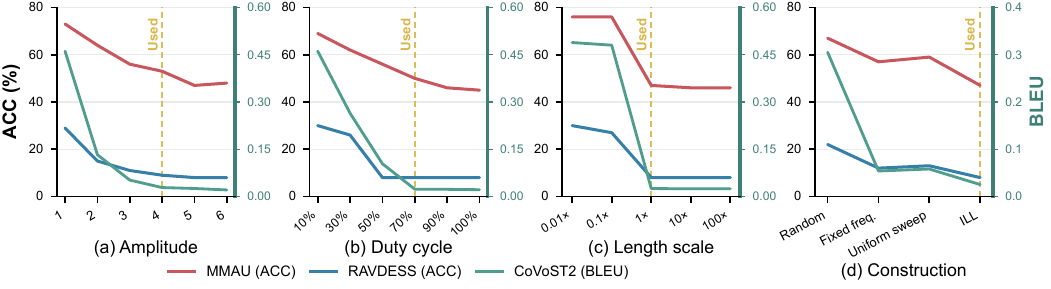} % Reduce the figure size so that it is slightly narrower than the column.
\caption{ILL component analysis on Qwen3-Omni. Panels vary (a) amplitude, (b) duty cycle, (c) segment length, and (d) state-sequence construction. Dashed lines mark the configuration used in the main experiments.}
\label{Fig:fig_ablation_component_analysis}
% \vspace{-5pt}
\end{figure*}

%% file: Table/table_attack_time.tex
% ==================== 导言区需要 ====================
% \usepackage{booktabs}
% \usepackage{adjustbox}
% \usepackage[table]{xcolor}
% \usepackage{colortbl}

% ==================== 正文表格 ====================
\begin{table}[t]
\centering
\caption{Corpus-median attention-derived active-interval duration $L_{\mathrm{on}}$ in seconds across datasets and LALMs.}
\label{tab:attention_interval_duration}

\footnotesize
\setlength{\tabcolsep}{4.5pt}
\renewcommand{\arraystretch}{1.15}
\arrayrulecolor{black}

\begin{adjustbox}{max width=\columnwidth,center}
\begin{tabular}{@{}l|cccc@{}}
\toprule

% 蓝灰色表头
\rowcolor[HTML]{DDE5ED}
\textbf{Dataset}\quad\quad\quad
& \quad\textbf{Qwen2.5}\quad\quad
& \quad\textbf{Qwen3}\quad\quad
& \quad\textbf{MiniCPM}\quad\quad
& \quad\textbf{StepAudio}\quad\quad \\

\midrule

% 第一行：白色
LibriSpeech
& 4.60
& 4.31
& 4.10
& 4.24 \\

% 第二行：浅灰色
\rowcolor[HTML]{F5F7F9}
CoVoST2
& 5.07
& 4.83
& 4.96
& 4.69 \\

% 第三行：白色
MMAU
& 4.56
& 4.37
& 4.47
& 4.55 \\

% 第四行：浅灰色
\rowcolor[HTML]{F5F7F9}
RAVDESS
& 3.69
& 3.69
& 3.69
& 3.67 \\

\bottomrule
\end{tabular}
\end{adjustbox}

% \vspace{-5pt}
\end{table}

%% file: sections/limitations.tex
\section{Limitations}
\label{sec:limitations}
For responsible evaluation, we do not deploy ILL in uncontrolled scenarios, where red teaming could affect unintended devices or users and cause potentially irreversible consequences. Our experiments instead simulate microphone reception and do not reproduce the complete acoustic path from a loudspeaker through a real-world acoustic environment to a microphone, leaving effects specific to devices and environments for controlled physical evaluation. Finally, our analysis characterizes changes in attention, representations, and output confidence rather than establishing a complete causal mechanism; this work prioritizes identifying the risk and demonstrating the effectiveness of a preliminary defense, while deeper mechanistic analysis remains future work.

%% file: Table/table_analyse.tex
% ==================== 正文表格 ====================
\begin{table}[t]
\centering
\caption{Internal attention, representation, and confidence statistics under clean, attack, and defense settings.}
\label{tab:internal_representation}

\footnotesize
\setlength{\tabcolsep}{3.2pt}
\renewcommand{\arraystretch}{1.15}
\arrayrulecolor{black}

\begin{adjustbox}{max width=\columnwidth,center}
\begin{tabular}{
  @{}
  l
  |ccc|ccc
  @{}
}
\toprule

% 第一行表头
\rowcolor[HTML]{DDE5ED}
\textbf{Dimension}\quad\quad\quad
& \multicolumn{3}{c|}{\textbf{MMAU}}
& \multicolumn{3}{c}{\textbf{RAVDESS}} \\

% 第二行表头
\rowcolor[HTML]{DDE5ED}
&
Clean
& Attack
& Defense
& Clean
& Attack
& Defense \\

\midrule

% 第一行：白色
Audio Attention Mass
& 0.041
& 0.037
& 0.039
& 0.077
& 0.053
& 0.064 \\

% 第二行：浅灰色
\rowcolor[HTML]{F5F7F9}
Cosine Similarity
& 1.000
& 0.865
& 0.958
& 1.000
& 0.630
& 0.955 \\

% 第三行：白色
Correct-Answer Probability
& 0.805
& 0.170
& 0.663
& 0.824
& 0.017
& 0.727 \\

% 第四行：浅灰色
\rowcolor[HTML]{F5F7F9}
Final-Answer Probability
& 0.805
& 0.636
& 0.732
& 0.824
& 0.777
& 0.811 \\

\bottomrule
\end{tabular}
\end{adjustbox}

% \vspace{-5pt}
\end{table}

%% file: sections/conclusion.tex
\section{Conclusion}
\label{sec:conclusion}
This work identifies inaudible low-frequency input as a practical hidden channel for disrupting audio understanding in LALMs and introduces \textit{Intermittent Low-Frequency Lockout} (ILL), a black box red teaming method that probes this risk with a fixed universal waveform. Across six LALMs and multiple speech tasks, ILL transfers without target-specific optimization, substantially degrades task performance, and limits spectral leakage into the measured audible band to 0.06--0.08\%. Its mean human audibility rating is 1.33, close to 1.17 for clean audio, while ablations show that its task-level effect depends on structured state construction rather than low-frequency energy alone. To mitigate this risk, we develop \textit{Distributional Requery Guard} (DRG), which detects low-frequency distribution shifts with F1 scores of 89.69--99.00\% under the Average training condition and uses conditional reacquisition to raise mean attacked accuracy from 28.5\% to 46.1\% when the second recording is clean, while preserving utility in most benign settings. Our findings establish low-frequency interference as a concrete robustness concern and provide an evaluation and mitigation basis for building audio-language systems whose behavior is better aligned with what users can perceive.

%% file: sections/appendix.tex
\section{Physical-Chain Feasibility and Scope}
\label{sec:appendix_physical_chain}

Our experiments control the amplitude of ILL at the waveform received by the model. To clarify what this receiver-side evaluation implies for an acoustic deployment, we separate \emph{reachability} from \emph{fidelity}. The former is physically feasible with an appropriate chain; the latter is neither required by our attack formulation nor guaranteed by commodity hardware.

\paragraph{End-to-end criterion.}
An emitted perturbation reaches the model only after passing through the source, the acoustic propagation environment, the microphone, the analog frontend, and the digital processing stages. The propagation environment may be indoor or outdoor and may include free-field propagation, reflections, obstacles, weather, and ambient noise. Each stage may attenuate or distort the 5--20~Hz component. The component remains available to the LALM when it is not completely removed by this chain and its received level stays above the combined acoustic, electronic, and quantization noise floor. It therefore need not have the same gain or phase as the speech band. The parameter $\beta$ in Section~\ref{sec:method_audio_semantic_transfer} represents this \emph{received} amplitude, rather than a device-independent loudspeaker setting.

\paragraph{Generation and acoustic propagation.}
The 20~Hz boundary is a convention tied to human audibility, not a propagation cutoff in air. Controlled sources have generated and transmitted signals well below this boundary. Park and Robertson constructed a compact rotary source and measured coherent infrasound at a 3.8~km stand-off distance, with received signal-to-noise ratios of 5--15~dB~\citep{park2009portable}. Asmar et al.\ subsequently used a portable rotary subwoofer to generate both tones and broadband pressure waves across 0.01--20~Hz in a non-isolated room~\citep{asmar2018digital}. Together, these outdoor and indoor measurements establish that the full ILL band can exist as an airborne pressure signal and propagate across diverse acoustic environments. They do not imply that an arbitrary laptop or phone loudspeaker can produce the same received level: conventional transduction technologies face mass-dominated radiation impedance at very low frequencies~\citep{park2009portable}. A physical realization of our threat model consequently requires a low-frequency-capable source and sufficient source level at the target location.

\paragraph{Microphone and capture response.}
Low-frequency reception is also experimentally established, including with compact transducers. Jeng et al.\ showed that a small electret condenser microphone connected to a notebook or desktop computer captured signals from 0.5 to 20~Hz in a controlled acoustic coupling arrangement~\citep{jeng2011response}. Asmar et al.\ characterized a digital smartphone microphone against a reference sensor over 0.01--20~Hz and recovered a parameterized infrasonic response in a non-isolated environment~\citep{asmar2018digital}. More recent field measurements using unmodified Samsung smartphone microphones recorded explosion energy concentrated near 10 and 20~Hz, while also showing strong low-frequency roll-off and phase distortion relative to a calibrated infrasound microphone~\citep{takazawa2024comparison}. Together, these studies support \emph{detectability}, but not a universal flat response: microphone ports, AC coupling, analog high-pass filters, automatic gain control, and manufacturer-specific processing can change the received amplitude and phase.

\section{Computational Complexity of DRG}
\label{sec:appendix_drg_complexity}

The quantities $U_x$ and $V$ retain their definitions from Section~\ref{sec:method_audio_semantic_transfer}, and all other notation follows Section~\ref{sec:method_distributional_requery_guard}.

\paragraph{Offline fitting.}
For each $x_{\mathrm{def}}\in\mathcal{D}_{\mathrm{def}}$, an FFT-based short-time Fourier transform costs $\mathcal{O}(U_{x_{\mathrm{def}}}V\log V)$, summing magnitudes over time costs $\mathcal{O}(U_{x_{\mathrm{def}}}V)$, and $\ell_1$ normalization costs $\mathcal{O}(V)$. Constructing all descriptors therefore costs $\mathcal{O}\!\left(\sum_{x_{\mathrm{def}}\in\mathcal{D}_{\mathrm{def}}}U_{x_{\mathrm{def}}}V\log V\right)$ and storing them costs $\mathcal{O}(|\mathcal{D}_{\mathrm{def}}|V)$. Because the number of clusters is fixed at two, each K-means assignment pass evaluates two $V$-dimensional distances for every descriptor, and centroid recomputation traverses the same descriptors; both operations cost $\mathcal{O}(|\mathcal{D}_{\mathrm{def}}|V)$ per fitting iteration. Labeling the two clusters with Equation~\ref{eq:defense_cluster_label} requires at most one additional scan of the descriptors and has the same linear bound. All of these costs are incurred offline.

\paragraph{Deployment.}
For $x^{(1)}$, descriptor extraction costs $\mathcal{O}(U_{x^{(1)}}V\log V)$ for the short-time Fourier transform and $\mathcal{O}(U_{x^{(1)}}V)$ for temporal aggregation. Matching $\mathbf{d}_x$ against $\boldsymbol{\mu}_1$ and $\boldsymbol{\mu}_2$ then performs exactly two Euclidean distance evaluations, giving $\mathcal{O}(V)$ detection-head time and $\mathcal{O}(V)$ storage for the descriptor and two fixed centroids. This step requires neither a neural network forward pass nor a target-LALM query. Only inputs assigned to $k_{\mathrm{jam}}$ trigger acquisition of $x^{(2)}$ and the subsequent target-LALM comparison; this conditional recovery cost is separate from the detector itself.

\section{Dataset Sampling and Aggregation}
\label{sec:appendix_data}

We sample 100 examples from each evaluated dataset or dataset subset. For MMAU, the speech, sound, and music subsets are evaluated separately, with 100 examples sampled from each subset. When a result is reported under the single label ``MMAU,'' it aggregates the 300 examples from these three subsets.

The same aggregation convention applies to Perceptual Noise Labels (PNL). PNL contains three interference types: Natural, Mechanical, and Human. When only a single ``PNL'' result is reported, it aggregates 300 evaluated examples, comprising 100 examples for each interference type.

For the Average training condition of the defense detector, training examples are randomly drawn from different datasets. The resulting training set contains 100 clean examples and 100 examples containing noise interference.

\section{Attack Baselines and Common Amplitude}
\label{sec:appendix_attack_baselines}

The attack baselines represent four distinct forms of audio interference. Gaussian uses randomly generated noise and serves as an unstructured random-noise baseline. PNL uses recorded environmental interference from its Natural, Machine, and Human categories and represents noise that may occur in everyday acoustic environments. Audio-Adv. is a white-box adversarial method that optimizes the audio input using access to the attacked model. Whisper is a transferable attack designed to remain effective beyond the model or setting used to construct it.

\paragraph{Common RMS budget.}
We compare attacks under a common waveform-domain energy budget, rather than by matching an unnormalised peak value. For each method $m$, let $p_m\in\mathbb{R}^{N}$ denote the complete perturbation aligned to an $N$-sample source utterance (including the off intervals of ILL). We remove any DC component and apply the standard root-mean-square (RMS) normalisation
\begin{equation}
    \begin{aligned}
        \bar{p}_m
        &=p_m-\frac{1}{N}\sum_{n=1}^{N}p_m[n],\\
        \delta_m
        &=\beta\frac{\bar{p}_m}
        {\sqrt{N^{-1}\sum_{n=1}^{N}\bar{p}_m[n]^2}},
        \qquad \beta=4.
    \end{aligned}
    \label{eq:common_rms_budget}
\end{equation}
Thus, before mixing, every non-clean method has the same RMS, $\operatorname{RMS}(\delta_m)=4$, over the full input duration. This protocol applies equally to Gaussian noise, PNL, Audio-Adv., Whisper, and the complete intermittent ILL waveform; it therefore accounts for ILL's envelope and inactive intervals instead of equating only its instantaneous peak with that of broadband noise. The model receives the digital composite $x+\delta_m$ used in the corresponding attack evaluation. 

For PNL, if an interference recording is shorter than the corresponding source audio, the interference recording is replayed cyclically until it covers the full duration of the source audio before applying Equation~\ref{eq:common_rms_budget}.

\section{Expanded MMAU Attack Results}
\label{sec:appendix_expanded_mmau}

The main-text result in Table~\ref{tab:attack_accuracy_mini} averages the speech, sound, and music categories of MMAU and further averages the Natural, Mechanical, and Human categories of PNL. Tables~\ref{tab:mmau_expanded_qwen25}--\ref{tab:mmau_expanded_gemini} expand both dimensions for every evaluated model. Model, task, and attack names follow the conventions used in the main text. As in Table~\ref{tab:attack_accuracy_mini}, lower attacked accuracy indicates a stronger disruption; subscripts show the accuracy change from Clean in percentage points, with gray for drops of at most 10 points, red for larger drops, and green for gains.

\paragraph{Consistency and task-level analysis.}
Re-averaging the three MMAU rows and the three PNL columns reproduces every value in Table~\ref{tab:attack_accuracy_mini} after rounding to one decimal place. Across the 18 model--task cells, ILL lowers accuracy relative to Clean in all 18 cases, with a mean reduction of 13.1 percentage points and a median reduction of 11.0 points (one-sided paired Wilcoxon signed-rank test, $p=3.81\times10^{-6}$). The mean reduction is largest on MMAU$_{\mathrm{Speech}}$ (21.5 points), followed by MMAU$_{\mathrm{Sound}}$ (9.3 points) and MMAU$_{\mathrm{Music}}$ (8.5 points). Thus, the aggregate MMAU result is not driven by a single model, although the magnitude varies substantially by task.

\paragraph{Paired nonparametric checks.}
We treat each model--task combination as one repeated-measures block and compare the seven attacks over the same 18 blocks. A tie-corrected Friedman test indicates that the attack outcomes are not exchangeable across methods. We then compare ILL with each baseline using one-sided paired Wilcoxon signed-rank tests and apply Holm correction across the six comparisons. Table~\ref{tab:mmau_expanded_tests} reports the post-hoc results. Positive mean differences denote higher baseline accuracy and therefore stronger disruption by ILL. ILL produces lower accuracy than Audio-Adv. in 17 of 18 cells and than Whisper in all 18 cells; both differences remain significant after correction. In contrast, the differences from Gaussian and the three individual PNL categories are not significant after correction, supporting a bounded conclusion of comparable task-level disruption rather than uniform dominance over these noise baselines.

\input{Table/table_attack_mmau_expanded}

\section{Human Audibility Evaluation}
\label{sec:appendix_human_evaluation}

The human audibility evaluation includes 112 distinct participants. Participants were publicly recruited student volunteers enrolled in undergraduate or postgraduate education, and no compensation was provided.

\paragraph{Stimuli and playback.}
Participants evaluated the same digitally mixed composite waveform $x+\delta_m$ that was provided to the model, rather than an isolated perturbation. Consequently, the listening stimulus preserves the interaction between the source recording and the interference. All non-clean stimuli use the common RMS budget in Equation~\ref{eq:common_rms_budget}; the methods therefore have the same waveform-domain RMS and energy per sample at playback. Each participant listened through one of two conventional consumer playback modes: headphones or direct loudspeaker playback. The playback volume was held fixed within a session, and participants did not adjust the audio level between trials. Including both modes prevents the conclusion from depending on one particular output path.

We collect subjective judgments using a seven-point rating scale, following a commonly studied Likert-type response format~\citep{Dawes_2008}. A rating of 1 indicates that the participant cannot perceive any noise, whereas a rating of 7 indicates that the noise is highly noticeable. Participants select one score from 1 to 7 to indicate the perceived audibility of the noise in each evaluated audio sample. The interpretation of each score is provided in Table~\ref{tab:human_audibility_scale}.

\begin{table}[H]
\centering
\caption{Interpretation of the seven-point human audibility scale.}
\label{tab:human_audibility_scale}
\small
\begin{tabular}{@{}cl@{}}
\toprule
\textbf{Score} & \textbf{Description} \\
\midrule
1 & Completely imperceptible \\
2 & Almost imperceptible \\
3 & Barely noticeable \\
4 & Slightly noticeable \\
5 & Moderately noticeable \\
6 & Clearly noticeable \\
7 & Highly noticeable \\
\bottomrule
\end{tabular}
\end{table}

\section{Ethics Statement}
\label{sec:appendix_ethics}

This work investigates a dual-use audio security risk with the goal of supporting controlled red teaming and mitigation. The study did not deploy ILL in public or otherwise uncontrolled acoustic environments. Attack evaluations used prerecorded benchmark audio and simulated the waveform received by the target microphone; they did not expose bystanders to an emitted low-frequency signal or connect model errors to decisions in consequential domains. The evaluated outcomes were limited to task-level errors in audio question answering, speech recognition, speech translation, and emotion classification. The human audibility evaluation likewise used prerecorded composite stimuli played through conventional headphones or loudspeakers. Participation was voluntary and uncompensated, and the manuscript reports only aggregate response statistics.

We acknowledge that the attack method could be misused to interfere with audio-enabled systems. To limit this risk, the study does not validate a field deployment or provide device-specific settings for emitted sound pressure, source placement, propagation, or microphone calibration. We also develop and evaluate DRG as a lightweight mitigation and explicitly document the dependence of physical feasibility and recovery on hardware and acoustic conditions. These measures do not eliminate the method's dual-use potential, but the controlled study involved no deployment in user-facing or consequential settings, and we observed no substantive harm to participants, bystanders, services, or physical systems.

%% file: Table/table_attack_mmau_expanded.tex
\begin{table*}[t]
\centering
\caption{Expanded MMAU attack results for Qwen2.5 (accuracy, \%). Subscripts give changes from Clean in percentage points; boldface marks ILL when it is lowest or tied for lowest.}
\label{tab:mmau_expanded_qwen25}
\scriptsize
\setlength{\tabcolsep}{3pt}
\renewcommand{\arraystretch}{1.12}
\begin{adjustbox}{max width=\textwidth,center}
\begin{tabular}{@{}lcccccccc@{}}
\toprule
\rowcolor[HTML]{DDE5ED}
\textbf{Task} & \textbf{Clean} & \textbf{Gaussian} & \textbf{PNL (Natural)} & \textbf{PNL (Mechanical)} & \textbf{PNL (Human)} & \textbf{Audio-Adv.} & \textbf{Whisper} & \textbf{ILL (Our)} \\
\midrule
MMAU$_{\mathrm{Speech}}$ & 69\% & \score{53}{16} & \score{53}{16} & \score{56}{13} & \score{56}{13} & \score{67}{2} & \score{72}{-3} & \bestscore{53}{16} \\
\rowcolor[HTML]{F5F7F9}
MMAU$_{\mathrm{Sound}}$ & 83\% & \score{71}{12} & \score{70}{13} & \score{69}{14} & \score{70}{13} & \score{74}{9} & \score{79}{4} & \score{75}{8} \\
MMAU$_{\mathrm{Music}}$ & 68\% & \score{60}{8} & \score{59}{9} & \score{58}{10} & \score{57}{11} & \score{67}{1} & \score{66}{2} & \bestscore{56}{12} \\
\bottomrule
\end{tabular}
\end{adjustbox}
\end{table*}

\begin{table*}[t]
\centering
\caption{Expanded MMAU attack results for Qwen3 (accuracy, \%). Subscripts give changes from Clean in percentage points; boldface marks ILL when it is lowest or tied for lowest.}
\label{tab:mmau_expanded_qwen3}
\scriptsize
\setlength{\tabcolsep}{3pt}
\renewcommand{\arraystretch}{1.12}
\begin{adjustbox}{max width=\textwidth,center}
\begin{tabular}{@{}lcccccccc@{}}
\toprule
\rowcolor[HTML]{DDE5ED}
\textbf{Task} & \textbf{Clean} & \textbf{Gaussian} & \textbf{PNL (Natural)} & \textbf{PNL (Mechanical)} & \textbf{PNL (Human)} & \textbf{Audio-Adv.} & \textbf{Whisper} & \textbf{ILL (Our)} \\
\midrule
MMAU$_{\mathrm{Speech}}$ & 75\% & \score{50}{25} & \score{55}{20} & \score{56}{19} & \score{61}{14} & \score{67}{8} & \score{73}{2} & \bestscore{47}{28} \\
\rowcolor[HTML]{F5F7F9}
MMAU$_{\mathrm{Sound}}$ & 81\% & \score{65}{16} & \score{68}{13} & \score{67}{14} & \score{70}{11} & \score{78}{3} & \score{78}{3} & \score{74}{7} \\
MMAU$_{\mathrm{Music}}$ & 68\% & \score{57}{11} & \score{57}{11} & \score{56}{12} & \score{56}{12} & \score{63}{5} & \score{64}{4} & \score{57}{11} \\
\bottomrule
\end{tabular}
\end{adjustbox}
\end{table*}

\begin{table*}[t]
\centering
\caption{Expanded MMAU attack results for MiniCPM (accuracy, \%). Subscripts give changes from Clean in percentage points; boldface marks ILL when it is lowest or tied for lowest.}
\label{tab:mmau_expanded_minicpm}
\scriptsize
\setlength{\tabcolsep}{3pt}
\renewcommand{\arraystretch}{1.12}
\begin{adjustbox}{max width=\textwidth,center}
\begin{tabular}{@{}lcccccccc@{}}
\toprule
\rowcolor[HTML]{DDE5ED}
\textbf{Task} & \textbf{Clean} & \textbf{Gaussian} & \textbf{PNL (Natural)} & \textbf{PNL (Mechanical)} & \textbf{PNL (Human)} & \textbf{Audio-Adv.} & \textbf{Whisper} & \textbf{ILL (Our)} \\
\midrule
MMAU$_{\mathrm{Speech}}$ & 72\% & \score{54}{18} & \score{55}{17} & \score{51}{21} & \score{55}{17} & \score{68}{4} & \score{68}{4} & \bestscore{49}{23} \\
\rowcolor[HTML]{F5F7F9}
MMAU$_{\mathrm{Sound}}$ & 77\% & \score{69}{8} & \score{70}{7} & \score{66}{11} & \score{69}{8} & \score{75}{2} & \score{79}{-2} & \score{72}{5} \\
MMAU$_{\mathrm{Music}}$ & 66\% & \score{60}{6} & \score{60}{6} & \score{61}{5} & \score{61}{5} & \score{65}{1} & \score{66}{0} & \bestscore{60}{6} \\
\bottomrule
\end{tabular}
\end{adjustbox}
\end{table*}

\begin{table*}[t]
\centering
\caption{Expanded MMAU attack results for StepAudio2 (accuracy, \%). Subscripts give changes from Clean in percentage points; boldface marks ILL when it is lowest or tied for lowest.}
\label{tab:mmau_expanded_stepaudio2}
\scriptsize
\setlength{\tabcolsep}{3pt}
\renewcommand{\arraystretch}{1.12}
\begin{adjustbox}{max width=\textwidth,center}
\begin{tabular}{@{}lcccccccc@{}}
\toprule
\rowcolor[HTML]{DDE5ED}
\textbf{Task} & \textbf{Clean} & \textbf{Gaussian} & \textbf{PNL (Natural)} & \textbf{PNL (Mechanical)} & \textbf{PNL (Human)} & \textbf{Audio-Adv.} & \textbf{Whisper} & \textbf{ILL (Our)} \\
\midrule
MMAU$_{\mathrm{Speech}}$ & 66\% & \score{50}{16} & \score{51}{15} & \score{51}{15} & \score{51}{15} & \score{61}{5} & \score{64}{2} & \bestscore{44}{22} \\
\rowcolor[HTML]{F5F7F9}
MMAU$_{\mathrm{Sound}}$ & 76\% & \score{66}{10} & \score{65}{11} & \score{60}{16} & \score{65}{11} & \score{72}{4} & \score{74}{2} & \score{66}{10} \\
MMAU$_{\mathrm{Music}}$ & 68\% & \score{66}{2} & \score{62}{6} & \score{65}{3} & \score{56}{12} & \score{69}{-1} & \score{66}{2} & \score{59}{9} \\
\bottomrule
\end{tabular}
\end{adjustbox}
\end{table*}

\begin{table*}[t]
\centering
\caption{Expanded MMAU attack results for GPT (accuracy, \%). Subscripts give changes from Clean in percentage points; boldface marks ILL when it is lowest or tied for lowest.}
\label{tab:mmau_expanded_gpt}
\scriptsize
\setlength{\tabcolsep}{3pt}
\renewcommand{\arraystretch}{1.12}
\begin{adjustbox}{max width=\textwidth,center}
\begin{tabular}{@{}lcccccccc@{}}
\toprule
\rowcolor[HTML]{DDE5ED}
\textbf{Task} & \textbf{Clean} & \textbf{Gaussian} & \textbf{PNL (Natural)} & \textbf{PNL (Mechanical)} & \textbf{PNL (Human)} & \textbf{Audio-Adv.} & \textbf{Whisper} & \textbf{ILL (Our)} \\
\midrule
MMAU$_{\mathrm{Speech}}$ & 64\% & \score{47}{17} & \score{54}{10} & \score{60}{4} & \score{53}{11} & \score{59}{5} & \score{62}{2} & \score{50}{14} \\
\rowcolor[HTML]{F5F7F9}
MMAU$_{\mathrm{Sound}}$ & 67\% & \score{52}{15} & \score{53}{14} & \score{58}{9} & \score{58}{9} & \score{57}{10} & \score{64}{3} & \bestscore{52}{15} \\
MMAU$_{\mathrm{Music}}$ & 52\% & \score{54}{-2} & \score{51}{1} & \score{50}{2} & \score{49}{3} & \score{59}{-7} & \score{56}{-4} & \bestscore{48}{4} \\
\bottomrule
\end{tabular}
\end{adjustbox}
\end{table*}

\begin{table*}[t]
\centering
\caption{Expanded MMAU attack results for Gemini (accuracy, \%). Subscripts give changes from Clean in percentage points; boldface marks ILL when it is lowest or tied for lowest.}
\label{tab:mmau_expanded_gemini}
\scriptsize
\setlength{\tabcolsep}{3pt}
\renewcommand{\arraystretch}{1.12}
\begin{adjustbox}{max width=\textwidth,center}
\begin{tabular}{@{}lcccccccc@{}}
\toprule
\rowcolor[HTML]{DDE5ED}
\textbf{Task} & \textbf{Clean} & \textbf{Gaussian} & \textbf{PNL (Natural)} & \textbf{PNL (Mechanical)} & \textbf{PNL (Human)} & \textbf{Audio-Adv.} & \textbf{Whisper} & \textbf{ILL (Our)} \\
\midrule
MMAU$_{\mathrm{Speech}}$ & 79\% & \score{54}{25} & \score{60}{19} & \score{63}{16} & \score{63}{16} & \score{75}{4} & \score{74}{5} & \bestscore{53}{26} \\
\rowcolor[HTML]{F5F7F9}
MMAU$_{\mathrm{Sound}}$ & 76\% & \score{68}{8} & \score{54}{22} & \score{53}{23} & \score{54}{22} & \score{73}{3} & \score{70}{6} & \score{65}{11} \\
MMAU$_{\mathrm{Music}}$ & 66\% & \score{63}{3} & \score{58}{8} & \score{63}{3} & \score{60}{6} & \score{65}{1} & \score{65}{1} & \bestscore{57}{9} \\
\bottomrule
\end{tabular}
\end{adjustbox}
\end{table*}

\begin{table*}[t]
\centering
\caption{Paired comparisons between ILL and the six attack baselines across the 18 model--task cells.}
\label{tab:mmau_expanded_tests}
\small
\setlength{\tabcolsep}{6pt}
\renewcommand{\arraystretch}{1.12}
\begin{tabular}{@{}lccc@{}}
\toprule
\rowcolor[HTML]{DDE5ED}
\textbf{Baseline} & \textbf{Mean accuracy difference (pp)} & \textbf{ILL lower/tied/higher} & \textbf{Holm-adjusted $p$} \\
\midrule
Gaussian & 1.2 & 9/5/4 & .422 \\
\rowcolor[HTML]{F5F7F9}
PNL (Natural) & 1.0 & 10/3/5 & .422 \\
PNL (Mechanical) & 1.4 & 12/0/6 & .422 \\
\rowcolor[HTML]{F5F7F9}
PNL (Human) & 1.5 & 11/0/7 & .422 \\
Audio-Adv. & 9.8 & 17/0/1 & $3.81\times10^{-5}$ \\
\rowcolor[HTML]{F5F7F9}
Whisper & 11.3 & 18/0/0 & $2.29\times10^{-5}$ \\
\bottomrule
\end{tabular}
\end{table*}

%% file: main.bbl
\begin{thebibliography}{45}
\providecommand{\natexlab}[1]{#1}

\bibitem[{Allen and Rabiner(1977)}]{allen1977unified}
Allen, J.~B.; and Rabiner, L.~R. 1977.
\newblock A unified approach to short-time Fourier analysis and synthesis.
\newblock \emph{Proceedings of the IEEE}, 65(11): 1558--1564.

\bibitem[{Asmar et~al.(2018)Asmar, Garc{\'e}s, Hart, and Williams}]{asmar2018digital}
Asmar, K.; Garc{\'e}s, M.; Hart, D.; and Williams, B. 2018.
\newblock Digital acoustic sensor performance across the infrasound range in non-isolated conditions.
\newblock \emph{The Journal of the Acoustical Society of America}, 144(5): 3036--3045.

\bibitem[{Carlini et~al.(2016)Carlini, Mishra, Vaidya, Zhang, Sherr, Shields, Wagner, and Zhou}]{carlini2016hidden}
Carlini, N.; Mishra, P.; Vaidya, T.; Zhang, Y.; Sherr, M.; Shields, C.; Wagner, D.; and Zhou, W. 2016.
\newblock Hidden voice commands.
\newblock In \emph{25th USENIX security symposium (USENIX security 16)}, 513--530.

\bibitem[{Carlini and Wagner(2018)}]{carlini2018audio}
Carlini, N.; and Wagner, D. 2018.
\newblock Audio adversarial examples: Targeted attacks on speech-to-text.
\newblock In \emph{2018 IEEE security and privacy workshops (SPW)}, 1--7. IEEE.

\bibitem[{Chen et~al.(2026)Chen, Song, Zhao, Jia, Liu, Qiao, Zhang, Tu, Yang, and Du}]{chen2026audiojailbreak}
Chen, G.; Song, F.; Zhao, Z.; Jia, X.; Liu, Y.; Qiao, Y.; Zhang, W.; Tu, W.; Yang, Y.; and Du, B. 2026.
\newblock Audiojailbreak: Jailbreak attacks against end-to-end large audio-language models.
\newblock \emph{IEEE Transactions on Dependable and Secure Computing}.

\bibitem[{Chu et~al.(2024)Chu, Xu, Yang, Wei, Wei, Guo, Leng, Lv, He, Lin et~al.}]{chu2024qwen2}
Chu, Y.; Xu, J.; Yang, Q.; Wei, H.; Wei, X.; Guo, Z.; Leng, Y.; Lv, Y.; He, J.; Lin, J.; et~al. 2024.
\newblock Qwen2-audio technical report.
\newblock \emph{arXiv preprint arXiv:2407.10759}.

\bibitem[{Chu et~al.(2023)Chu, Xu, Zhou, Yang, Zhang, Yan, Zhou, and Zhou}]{chu2023qwen}
Chu, Y.; Xu, J.; Zhou, X.; Yang, Q.; Zhang, S.; Yan, Z.; Zhou, C.; and Zhou, J. 2023.
\newblock Qwen-audio: Advancing universal audio understanding via unified large-scale audio-language models.
\newblock \emph{arXiv preprint arXiv:2311.07919}.

\bibitem[{Cui et~al.(2026)Cui, Xu, Wang, Yu, Sun, Xu, Wang, He, Ma, Cai et~al.}]{cui2026minicpm}
Cui, J.; Xu, B.; Wang, C.; Yu, T.; Sun, W.; Xu, Y.; Wang, T.; He, Z.; Ma, W.; Cai, T.; et~al. 2026.
\newblock Minicpm-o 4.5: Towards real-time full-duplex omni-modal interaction.
\newblock \emph{arXiv preprint arXiv:2604.27393}.

\bibitem[{Dawes(2008)}]{Dawes_2008}
Dawes, J. 2008.
\newblock Do data characteristics change according to the number of scale points used? An experiment using 5-point, 7-point and 10-point scales.
\newblock \emph{International journal of market research}, 50(1): 61--104.

\bibitem[{Djanibekov et~al.(2025)Djanibekov, Mukhituly, Inui, Aldarmaki, and Lukas}]{djanibekov2025spirit}
Djanibekov, A.; Mukhituly, N.; Inui, K.; Aldarmaki, H.; and Lukas, N. 2025.
\newblock Spirit: Patching speech language models against jailbreak attacks.
\newblock In \emph{Proceedings of the 2025 Conference on Empirical Methods in Natural Language Processing}, 14514--14531.

\bibitem[{Ephraim and Malah(1984)}]{ephraim1984speech}
Ephraim, Y.; and Malah, D. 1984.
\newblock Speech enhancement using a minimum-mean square error short-time spectral amplitude estimator.
\newblock \emph{IEEE Transactions on Acoustics Speech and Signal Processing}, 32(6): 1109--1121.

\bibitem[{Franceschi, Fawzi, and Fawzi(2018)}]{franceschi2018robustness}
Franceschi, J.-Y.; Fawzi, A.; and Fawzi, O. 2018.
\newblock Robustness of classifiers to uniform $\ell_p$ and Gaussian noise.
\newblock In \emph{International Conference on Artificial Intelligence and Statistics}, 1280--1288. PMLR.

\bibitem[{Germain, Chen, and Koltun(2018)}]{germain2018speech}
Germain, F.~G.; Chen, Q.; and Koltun, V. 2018.
\newblock Speech denoising with deep feature losses.
\newblock \emph{arXiv preprint arXiv:1806.10522}.

\bibitem[{Hou et~al.(2025)Hou, He, Zhou, Guo, Qiao, Zhang, and Jiang}]{hou2025evaluating}
Hou, G.; He, J.; Zhou, Y.; Guo, J.; Qiao, Y.; Zhang, R.; and Jiang, W. 2025.
\newblock Evaluating robustness of large audio language models to audio injection: An empirical study.
\newblock In \emph{Proceedings of the 2025 Conference on Empirical Methods in Natural Language Processing}, 25671--25687.

\bibitem[{Hsu et~al.(2025)Hsu, Lu, Chiang, and Lee}]{hsu2025reducing}
Hsu, T.-w.; Lu, K.-H.; Chiang, C.-H.; and Lee, H.-y. 2025.
\newblock Reducing object hallucination in large audio-language models via audio-aware decoding.
\newblock \emph{arXiv preprint arXiv:2506.07233}.

\bibitem[{Jeng, Yang, and Lee(2011)}]{jeng2011response}
Jeng, Y.-N.; Yang, T.-M.; and Lee, S.-Y. 2011.
\newblock Response identification in the extremely low frequency region of an electret condenser microphone.
\newblock \emph{Sensors}, 11(1): 623--637.

\bibitem[{Ko et~al.(2017)Ko, Peddinti, Povey, Seltzer, and Khudanpur}]{ko2017study}
Ko, T.; Peddinti, V.; Povey, D.; Seltzer, M.~L.; and Khudanpur, S. 2017.
\newblock A study on data augmentation of reverberant speech for robust speech recognition.
\newblock In \emph{2017 IEEE international conference on acoustics, speech and signal processing (ICASSP)}, 5220--5224. IEEE.

\bibitem[{Lin et~al.(2026)Lin, Yu, Luo, Zhang, Peng, Wang, Tang, Zhang, Yang, Zhou et~al.}]{lin2026hidden}
Lin, L.; Yu, M.; Luo, K.; Zhang, Y.; Peng, L.; Wang, D.; Tang, X.; Zhang, Y.; Yang, X.; Zhou, Z.; et~al. 2026.
\newblock Hidden in the noise: Unveiling backdoors in audio llms alignment through latent acoustic pattern triggers.
\newblock In \emph{Proceedings of the AAAI Conference on Artificial Intelligence}, volume~40, 32015--32023.

\bibitem[{Livingstone and Russo(2018)}]{livingstone2018ryerson}
Livingstone, S.~R.; and Russo, F.~A. 2018.
\newblock The Ryerson Audio-Visual Database of Emotional Speech and Song (RAVDESS): A dynamic, multimodal set of facial and vocal expressions in North American English.
\newblock \emph{PloS one}, 13(5): e0196391.

\bibitem[{Ma et~al.(2025)Ma, Qian, Raina, Gales, and Knill}]{ma2025universal}
Ma, R.; Qian, M.; Raina, V.; Gales, M.; and Knill, K. 2025.
\newblock Universal Acoustic Adversarial Attacks for Flexible Control of Speech-LLMs.
\newblock \emph{Association for Computational Linguistics (ACL)}.

\bibitem[{Panayotov et~al.(2015)Panayotov, Chen, Povey, and Khudanpur}]{panayotov2015librispeech}
Panayotov, V.; Chen, G.; Povey, D.; and Khudanpur, S. 2015.
\newblock Librispeech: an asr corpus based on public domain audio books.
\newblock In \emph{2015 IEEE international conference on acoustics, speech and signal processing (ICASSP)}, 5206--5210. IEEE.

\bibitem[{Papineni et~al.(2002)Papineni, Roukos, Ward, and Zhu}]{papineni2002bleu}
Papineni, K.; Roukos, S.; Ward, T.; and Zhu, W.-J. 2002.
\newblock Bleu: a method for automatic evaluation of machine translation.
\newblock In \emph{Proceedings of the 40th annual meeting of the Association for Computational Linguistics}, 311--318.

\bibitem[{Park and Robertson(2009)}]{park2009portable}
Park, J.; and Robertson, J. 2009.
\newblock A portable infrasound generator.
\newblock \emph{The Journal of the Acoustical Society of America}, 125(4): EL148--EL151.

\bibitem[{Peri et~al.(2024)Peri, Jayanthi, Ronanki, Bhatia, Mundnich, Dingliwal, Das, Hou, Huybrechts, Vishnubhotla et~al.}]{peri2024speechguard}
Peri, R.; Jayanthi, S.~M.; Ronanki, S.; Bhatia, A.; Mundnich, K.; Dingliwal, S.; Das, N.; Hou, Z.; Huybrechts, G.; Vishnubhotla, S.; et~al. 2024.
\newblock Speechguard: exploring the adversarial robustness of multi-modal large language models.
\newblock In \emph{Findings of the Association for Computational Linguistics: ACL 2024}, 10018--10035.

\bibitem[{Raina et~al.(2024)Raina, Ma, McGhee, Knill, and Gales}]{raina2024muting}
Raina, V.; Ma, R.; McGhee, C.; Knill, K.; and Gales, M. 2024.
\newblock Muting whisper: A universal acoustic adversarial attack on speech foundation models.
\newblock In \emph{Proceedings of the 2024 Conference on Empirical Methods in Natural Language Processing}, 7549--7565.

\bibitem[{Rubenstein et~al.(2023)Rubenstein, Asawaroengchai, Nguyen, Bapna, Borsos, Quitry, Chen, Badawy, Han, Kharitonov et~al.}]{rubenstein2023audiopalm}
Rubenstein, P.~K.; Asawaroengchai, C.; Nguyen, D.~D.; Bapna, A.; Borsos, Z.; Quitry, F. d.~C.; Chen, P.; Badawy, D.~E.; Han, W.; Kharitonov, E.; et~al. 2023.
\newblock Audiopalm: A large language model that can speak and listen.
\newblock \emph{arXiv preprint arXiv:2306.12925}.

\bibitem[{Sadasivan et~al.(2026)Sadasivan, Feizi, Mathews, and Wang}]{sadasivan2026attacker}
Sadasivan, V.~S.; Feizi, S.; Mathews, R.; and Wang, L. 2026.
\newblock Attacker’s noise can manipulate your audio-based llm in the real world.
\newblock In \emph{Proceedings of the 19th Conference of the European Chapter of the Association for Computational Linguistics (Volume 1: Long Papers)}, 1430--1440.

\bibitem[{Sakshi et~al.(2025)Sakshi, Tyagi, Kumar, Seth, Selvakumar, Nieto, Duraiswami, Ghosh, and Manocha}]{sakshi2025mmau}
Sakshi, S.; Tyagi, U.; Kumar, S.; Seth, A.; Selvakumar, R.; Nieto, O.; Duraiswami, R.; Ghosh, S.; and Manocha, D. 2025.
\newblock Mmau: A massive multi-task audio understanding and reasoning benchmark.
\newblock In \emph{International Conference on Learning Representations}, volume 2025, 84929--84964.

\bibitem[{Salt and Hullar(2010)}]{Salt_2010}
Salt, A.~N.; and Hullar, T.~E. 2010.
\newblock Responses of the ear to low frequency sounds, infrasound and wind turbines.
\newblock \emph{Hearing Research}, 268(1-2): 12--21.

\bibitem[{Song et~al.(2026)Song, Jiang, Cui, Li, Gao, Zhang, Xu, Wang, Ouyang, Chen et~al.}]{song2026audio}
Song, Z.; Jiang, Q.; Cui, M.; Li, M.; Gao, L.; Zhang, Z.; Xu, Z.; Wang, Y.; Ouyang, G.; Chen, Z.; et~al. 2026.
\newblock Audio jailbreak: An open comprehensive benchmark for jailbreaking large audio-language models.
\newblock In \emph{Proceedings of the 64th Annual Meeting of the Association for Computational Linguistics (Volume 1: Long Papers)}, 27294--27308.

\bibitem[{Takazawa et~al.(2024)Takazawa, Popenhagen, Ocampo~Giraldo, Cardenas, Hix, Thompson, Chichester, and Garc{\'e}s}]{takazawa2024comparison}
Takazawa, S.~K.; Popenhagen, S.; Ocampo~Giraldo, L.; Cardenas, E.; Hix, J.; Thompson, S.; Chichester, D.; and Garc{\'e}s, M. 2024.
\newblock A comparison of smartphone and infrasound microphone data from a fuel air explosive and a high explosive.
\newblock \emph{The Journal of the Acoustical Society of America}, 156(3): 1509--1523.

\bibitem[{Wang et~al.(2021)Wang, Wu, Gu, and Pino}]{wang2021covost}
Wang, C.; Wu, A.; Gu, J.; and Pino, J. 2021.
\newblock CoVoST 2 and Massively Multilingual Speech Translation.
\newblock In \emph{Proc. Interspeech 2021}, 2247--2251.

\bibitem[{Wu et~al.(2025)Wu, Yan, Hu, Yi, Feng, Tian, Shen, Yu, Zhang, Li et~al.}]{wu2025step}
Wu, B.; Yan, C.; Hu, C.; Yi, C.; Feng, C.; Tian, F.; Shen, F.; Yu, G.; Zhang, H.; Li, J.; et~al. 2025.
\newblock Step-audio 2 technical report.
\newblock \emph{arXiv preprint arXiv:2507.16632}.

\bibitem[{Xu et~al.(2025{\natexlab{a}})Xu, Guo, He, Hu, He, Bai, qin Chen, Wang, Fan, Dang, Zhang, Wang, Chu, and Lin}]{Xu2025Qwen25OmniTR}
Xu, J.; Guo, Z.; He, J.; Hu, H.; He, T.; Bai, S.; qin Chen, K.; Wang, J.; Fan, Y.; Dang, K.; Zhang, B.; Wang, X.; Chu, Y.; and Lin, J. 2025{\natexlab{a}}.
\newblock Qwen2.5-Omni Technical Report.
\newblock \emph{ArXiv}, abs/2503.20215.

\bibitem[{Xu et~al.(2025{\natexlab{b}})Xu, Guo, Hu, Chu, Wang, He, Wang, Shi, He, Zhu et~al.}]{xu2025qwen3}
Xu, J.; Guo, Z.; Hu, H.; Chu, Y.; Wang, X.; He, J.; Wang, Y.; Shi, X.; He, T.; Zhu, X.; et~al. 2025{\natexlab{b}}.
\newblock Qwen3-omni technical report.
\newblock \emph{arXiv preprint arXiv:2509.17765}.

\bibitem[{Yan et~al.(2020)Yan, Liu, Zhou, Guo, and Zhang}]{yan2020surfingattack}
Yan, Q.; Liu, K.; Zhou, Q.; Guo, H.; and Zhang, N. 2020.
\newblock Surfingattack: Interactive hidden attack on voice assistants using ultrasonic guided waves.
\newblock In \emph{Network and Distributed Systems Security (NDSS) Symposium}.

\bibitem[{Yang et~al.(2025{\natexlab{a}})Yang, Qu, Shareghi, and Haffari}]{yang2025audio}
Yang, H.; Qu, L.; Shareghi, E.; and Haffari, G. 2025{\natexlab{a}}.
\newblock Audio is the achilles’ heel: Red teaming audio large multimodal models.
\newblock In \emph{Proceedings of the 2025 Conference of the Nations of the Americas Chapter of the Association for Computational Linguistics: Human Language Technologies (Volume 1: Long Papers)}, 9292--9306.

\bibitem[{Yang et~al.(2025{\natexlab{b}})Yang, Li, Fang, Wei, and Chen}]{yang2025can}
Yang, W.; Li, Y.; Fang, M.; Wei, Y.; and Chen, L. 2025{\natexlab{b}}.
\newblock Who Can Withstand Chat-Audio Attacks? An Evaluation Benchmark for Large Audio-Language Models.
\newblock In \emph{Findings of the Association for Computational Linguistics: ACL 2025}, 17205--17220.

\bibitem[{Yang et~al.(2025{\natexlab{c}})Yang, Zhang, Han, Wang, Zhuang, Jin, Shao, Sun, and Zhang}]{yang2025speech}
Yang, Y.; Zhang, X.; Han, Z.; Wang, S.; Zhuang, J.; Jin, Z.; Shao, J.; Sun, G.; and Zhang, C. 2025{\natexlab{c}}.
\newblock Speech-Audio Compositional Attacks on Multimodal LLMs and Their Mitigation with SALMONN-Guard.
\newblock \emph{arXiv preprint arXiv:2511.10222}.

\bibitem[{Yin et~al.(2026)Yin, Xiao, Kwon, Dang, and Choi}]{yin2026focus}
Yin, H.; Xiao, Y.; Kwon, Y.; Dang, T.; and Choi, J.-W. 2026.
\newblock Focus Then Listen: An Empirical Study of Plug-and-Play Audio Enhancer for Noise-Robust Large Audio Language Models.
\newblock \emph{arXiv preprint arXiv:2603.04862}.

\bibitem[{Yu et~al.(2026)Yu, Jin, Yu, Zhuang, and Wang}]{yu2026now}
Yu, Y.; Jin, H.; Yu, Y.; Zhuang, J.; and Wang, H. 2026.
\newblock Now You Hear Me: Audio Narrative Attacks Against Large Audio-Language Models.
\newblock \emph{arXiv preprint arXiv:2601.23255}.

\bibitem[{Yuan et~al.(2018)Yuan, Chen, Zhao, Long, Liu, Chen, Zhang, Huang, Wang, and Gunter}]{yuan2018commandersong}
Yuan, X.; Chen, Y.; Zhao, Y.; Long, Y.; Liu, X.; Chen, K.; Zhang, S.; Huang, H.; Wang, X.; and Gunter, C.~A. 2018.
\newblock $\{$CommanderSong$\}$: A systematic approach for practical adversarial voice recognition.
\newblock In \emph{27th USENIX security symposium (USENIX security 18)}, 49--64.

\bibitem[{Zhang et~al.(2023)Zhang, Li, Zhang, Zhan, Wang, Zhou, and Qiu}]{zhang2023speechgpt}
Zhang, D.; Li, S.; Zhang, X.; Zhan, J.; Wang, P.; Zhou, Y.; and Qiu, X. 2023.
\newblock Speechgpt: Empowering large language models with intrinsic cross-modal conversational abilities.
\newblock In \emph{Findings of the Association for Computational Linguistics: EMNLP 2023}, 15757--15773.

\bibitem[{Zhang et~al.(2017)Zhang, Yan, Ji, Zhang, Zhang, and Xu}]{zhang2017dolphinattack}
Zhang, G.; Yan, C.; Ji, X.; Zhang, T.; Zhang, T.; and Xu, W. 2017.
\newblock Dolphinattack: Inaudible voice commands.
\newblock In \emph{Proceedings of the 2017 ACM SIGSAC conference on computer and communications security}, 103--117.

\bibitem[{Zhang et~al.(2026)Zhang, Tian, Zhang, Yan, Lin, Zhou, Sun, and Su}]{zhang2026see}
Zhang, Y.; Tian, J.; Zhang, Y.; Yan, S.; Lin, L.; Zhou, Z.; Sun, L.; and Su, S. 2026.
\newblock SEE: Signal Embedding Energy for Quantifying Noise Interference in Large Audio Language Models.
\newblock \emph{arXiv preprint arXiv:2601.07331}.

\end{thebibliography}
